\documentclass[sigconf, screen]{acmart}
\AtBeginDocument{%
  \providecommand\BibTeX{{%
    \normalfont B\kern-0.5em{\scshape i\kern-0.25em b}\kern-0.8em\TeX}}}
    
\usepackage{censor}
\usepackage{comment}

\usepackage{subfig}
\usepackage{graphicx}
\usepackage[htt]{hyphenat}
\newcommand*\circled[1]{\tikz[baseline=(char.base)]{
            \node[shape=circle,draw,inner sep=0.8pt] (char) {#1};}}
\newcommand{\ignore}[1]{}

\usepackage[noframe]{showframe}
\usepackage{framed}
\renewenvironment{shaded}{%
  \MakeFramed{\advance\hsize-\width \FrameRestore\FrameRestore}}%
 {\endMakeFramed}
\definecolor{shadecolor}{gray}{0.75}
    
\setcopyright{acmcopyright}
\copyrightyear{2021}
\acmYear{2021}
\acmDOI{10.1145/3485832.3485888}

\acmConference[ACSAC '21]{Annual Computer Security Applications Conference (ACSAC)}{December 6--December 10, 2021}{Online}
\acmBooktitle{ACSAC '21: Annual Computer Security Applications Conference (ACSAC),
  December 6--December 10, 2021, Online}
\acmISBN{978-1-4503-8579-4/21/12}

\begin{document}

\title{Characterizing Improper Input Validation Vulnerabilities of Mobile Crowdsourcing Services}


\author{Sojhal Ismail Khan}
\email{sojhalis@usc.edu}
\affiliation{%
  \department{Department of Computer Science}
  \institution{University of Southern California}
  \state{California}
  \country{USA}
}

\author{Dominika Woszczyk}
\email{d.woszczyk19@imperial.ac.uk}
\affiliation{%
  \department{Department of Computing}
  \institution{Imperial College London}
  \city{London}
  \country{UK}
}

\author{Chengzeng You}
\email{chengzeng.you19@imperial.ac.uk}
\affiliation{%
  \department{Department of Computing}
  \institution{Imperial College London}
  \city{London}
  \country{UK}
}

\author{Soteris Demetriou}
\email{s.demetriou@imperial.ac.uk}
\affiliation{%
  \department{Department of Computing}
  \institution{Imperial College London}
  \city{London}
  \country{UK}
}


\author{Muhammad Naveed}
\email{mnaveed@usc.edu}
\affiliation{%
  \department{Department of Computer Science}
  \institution{University of Southern California}
  \state{California}
  \country{USA}
}

\renewcommand{\shortauthors}{Khan, et al.}

\begin{abstract}

Mobile crowdsourcing services (MCS), enable fast and economical data acquisition at scale and find applications in a variety of domains. Prior work has shown that Foursquare and Waze (a location-based and a navigation MCS) are vulnerable to different kinds of data poisoning attacks. Such attacks can be upsetting and even dangerous especially when they are used to inject improper inputs to mislead users. However, to date, there is no comprehensive study on the extent of improper input validation (IIV) vulnerabilities and the feasibility of their exploits in MCSs across domains. In this work, we leverage the fact that MCS interface with their participants through mobile apps to design tools and new methodologies embodied in an end-to-end feedback-driven analysis framework which we use to study 10 popular and previously unexplored services in five different domains. Using our framework we send tens of thousands of API requests with automatically generated input values to characterize their IIV attack surface.
Alarmingly, we found that most of them (8/10) suffer from grave IIV vulnerabilities which allow an adversary to launch data poisoning attacks at scale: 7400 spoofed API requests were successful in faking online posts for robberies, gunshots, and other dangerous incidents, faking fitness activities with supernatural speeds and distances among many others. Lastly, we discuss easy to implement and deploy mitigation strategies which can greatly reduce the IIV attack surface and argue for their use as a necessary complementary measure working toward trustworthy mobile crowdsourcing services. 
\end{abstract}



\maketitle

\section{Introduction}
\label{sec:intro}
Mobile crowdsourcing services (MCSs) enable economical, rapid, and scalable data acquisition utilized for accurate information sharing for smart navigation and transportation (~\cite{GoogleMa92:online, Transit:online}); health and fitness recommendations~\cite{StravaSu80:online, FitbitOf86:online,MapMyRun46:online}; and price tracking~\cite{BasketSm94:online,FindTheN91:online} among others.
However,  location-based (Foursquare, Facebook Places~\cite{polakis2013man}) and navigation (Google Maps~\cite{SIMONWEC66:online}, Waze~\cite{wang2016defending}) MCSs have been shown to be susceptible to ad-hoc data poisoning attacks. For example, recently an individual showed how a real-world navigation service can be fooled to make wrong predictions on traffic density, allowing an adversary to redirect traffic~\cite{SIMONWEC66:online}. This experiment was performed manually by carrying 100 smartphones while walking up and down the target road. While this demonstrates both the feasibility and potential consequences of data poisoning attacks on MCS, the experiment is hard to replicate and systematically scale it to study fundamental issues enabling such exploits in other MCS. More systematic studies conducted by Polakis et al.~\cite{polakis2013man} and Wang et al.~\cite{wang2016defending}, while sound, they are specific to the characteristics of the target MCS and thus fall short in providing generalizing insights on the vulnerabilities of MCSs. 

We observe, that MCSs across application domains suffer from a common vulnerability, that of \textit{improper input validation} (\textit{IIV}) which can be exploited by an adversary to inject hazardous data or spread mis-information. To better understand the presence of IIV vulnerabilities in MCSs and to what extend they contribute to their exposure to improper input injection attacks,  we conduct a systematic analysis on 10 high-profile, previously unexplored MCSs across 5 different application domains. To perform our analysis we had to overcome two main challenges: firstly, the closed source nature of MCSs does not allow for trivial examination of their input validation mechanisms; and secondly, testing a large number of input values for different input types is impractical. To overcome the first challenge we present a feedback-driven analysis framework suitable for black-box analysis of MCSs. The framework leverages the observation that most MCSs interface with their participants through companion mobile apps and embodies a set of \textit{input injection components} for interacting with the remote service. The injection components  target three main avenues an adversary can exploit on companion mobile apps to inject improper inputs to the remote service: the sensor measurements used by the companion apps, their user interface inputs, and their network requests to the target service. The framework also uses \textit{feedback monitoring} strategies for facilitating the evaluation of each input injection. To overcome the second challenge, we introduce and integrate in the end-to-end analysis framework a set of \textit{input exploration strategies}. These generate values for various input types, supporting \textit{range and constraint} and \textit{semantic} input exploration.

We then design systematic experiments for all selected services and leverage our framework to characterize the extent of their exposure to improper input injections. Our analysis led to an array of alarming findings. We found that 8/10 services are vulnerable to such attacks.  Some of them do not perform any kind of input validation, accepting values which span across the expected input's domain \textit{range}. Some have some restrictions on the range but perform no \textit{semantic} validation. Even for the ones that do take measures to verify inputs we show that they are bypassable. Our findings are alarming: we were able to fake running activity speeds equivalent to 10x the speed of a commercial jet aircraft and run distances equivalent to running around the Earth 2007 times~(Section~\ref{sec:fitness}), reduce prices for commodity items down to 10\% their value and up to double their value (Section~\ref{sec:prices}), fake public bus rides (\ref{sec:transit}), inject fake places of interest in the middle of the ocean (Section~\ref{sec:location}), and fake reports for robberies, gunshots and other dangerous incidents in safety services (Section~\ref{sec:safety}). Demos of successful improper input injections can be found on our project's website~\cite{mcs_website}. To mitigate such issues, we discuss a set of backward-compatible and simple to implement input validation strategies which can reduce the attack surface of MCSs up to 99.58\%. These can complement and increase the efficiency and effectiveness of existing countermeasures such as reputation schemes, UI hints and majority voting.

\vspace{5pt}\noindent\textbf{Contributions.} Below we summarize our main contributions:

\vspace{3pt}\noindent$\bullet$ \textit{New Techniques.} We develop range and constraint, and semantic input exploration strategies for generating values for numeric inputs, GPS coordinates and social posts. We further design methods to simulate adversarial capabilities for spoofing network API requests, UI inputs and sensor inputs.

\vspace{3pt}\noindent$\bullet$ \textit{Framework for Analysis of IIVs in MCSs.} We introduce a feedback-driven framework which embodies input exploration and injection methods in tandem with feedback monitoring mechanisms to facilitate the analysis of IIV vulnerabilities in MCSs from the vantage point of their companion mobile apps.



\vspace{3pt}\noindent$\bullet$ \textit{New Findings.} We discovered and reported previously unknown vulnerabilities for 8 high-profile MCSs that can have grave consequences on their veracity and ensuing trustworthiness. 


\vspace{3pt}\noindent\textbf{Ethical Considerations.}
Even though this study was classified as IRB--exempt, we take various measures in our experiments to mitigate risk of affecting users or services. These include monitoring services' activity and focusing on regions and times to minimize the exposure of erroneous values to real users; and deleting or reverting values to their original state immediately after we verify their approval by the service.\ignore{To the best of our knowledge no user was affected.} See Appendix~\ref{sec:ethical_considerations_appendix} for further details.


\vspace{3pt}\noindent\textbf{Responsible Disclosure.} Affected services were contacted at least 3 months prior to this writing through their developer email address. Transit, MapMyRun and Fitbit responded with an automated email confirming receipt, but there was no follow-up. Strava requested further details which we provided. We also submitted reports to bug bounty programs of Fitbit and MapMyRun on bugcrowd.com. Fitbit responded stating that the issue is not applicable for a reward because they couldn't identify a security impact for their customers.


\vspace{3pt}\noindent\textbf{Paper Organization.}
The rest of the paper is organized as follows. In Section~\ref{sec:threats} we describe the problem and define our threat model. In Section~\ref{sec:framework} we present our analysis framework and main components. In Sections~\ref{sec:fitness}, \ref{sec:prices}, \ref{sec:transit}, \ref{sec:location}, \ref{sec:safety}, we elaborate on how we utilize our framework to characterize the attack surface of popular MCSs across the most important application domains. In Sections~\ref{sec:defense} and ~\ref{sec:related} we discuss mitigation strategies and related work respectively, and conclude the paper in Section~\ref{sec:conclusion}.

\section{Background and Threat Model}
\label{sec:threats}

\noindent\textbf{Improper Input Validation.} Mobile crowdsourcing services face the risk of data injection attacks. In such an attack, a malicious participant node injects an erroneous measurement in the global service aiming to force an error or deceive the users of the service. To achieve this, the adversary might target \textit{improper input validation} (IIV) vulnerabilities. Validation can be \textit{syntactic}, \textit{range and constraint} or \textit{semantic}. Lack of syntactic verification might cause crashes. For example, a service or its corresponding user-facing mobile application might expect the user to report a number. However, in light of insufficient input syntactic validation, an adversary might cause a service to crash by introducing a measurement of an unexpected input type. Fuzzing techniques are typically employed to uncover such reliability issues~\cite{chen2018iotfuzzer}. In this work, we focus mainly on range and constraint validation and semantic validation. Attacks leveraging these are harder to detect as at a first glance the reported values do not seem anomalous.

\vspace{3pt}\noindent$\bullet$\textit{ Range and Constraint validation.}
This step ensures that the input domain range is minimized to accept values meaningful to the context of the service. For example, a service expecting GPS coordinates as float values with lack of range validation might be poisoned with non-existing coordinates (e.g. longitude value greater than $180.0^{\circ}$). 

\vspace{3pt}\noindent$\bullet$\textit{ Semantic Validation.}
Semantic validation is used to validate the meaning of the input. An adversary exploiting the lack of semantic validation in GPS inputs might introduce float values that do correspond to valid GPS coordinates range but for an implausible location of the given point of interest.
\vspace{3pt}\noindent\textbf{Threat Model.}
We consider an adversary ($\mathcal{A}$) with access to the mobile app of the target ubiquitous crowdsourcing service. $\mathcal{A}$ can observe the traffic generated between the app and the remote service either by passive eavesdropping or active man in the middle attacks. $\mathcal{A}$ can also reverse engineer and analyze the mobile app interfacing with the service. Thus the adversary is in knowledge of the communication protocol and can leverage it to try to inject fake data into the service. However, the adversary has no access to the remote service and can only treat it as a black box.

\section{Analysis Framework}
\label{sec:framework}
\subsection{Overview}
To characterize the IIV attack surface of MCSs we need a systematic way of exploring the ability of $\mathcal{A}$ to launch successful attacks exploiting the lack of input validation in MCS. This is by no means a trivial endeavor.
One could perform improper input injection attempts by manually interacting with the services' companion mobile apps, but this would be both tiresome and impractical for a large number of test inputs. To make things worse, we do not have white box access to the MCSs and hence we cannot trivially determine if a service performs input validation.

\begin{figure}[!t]
    \centerline{\includegraphics[clip, trim=0.0cm 0.5cm 0.0cm 0cm, width=\columnwidth]{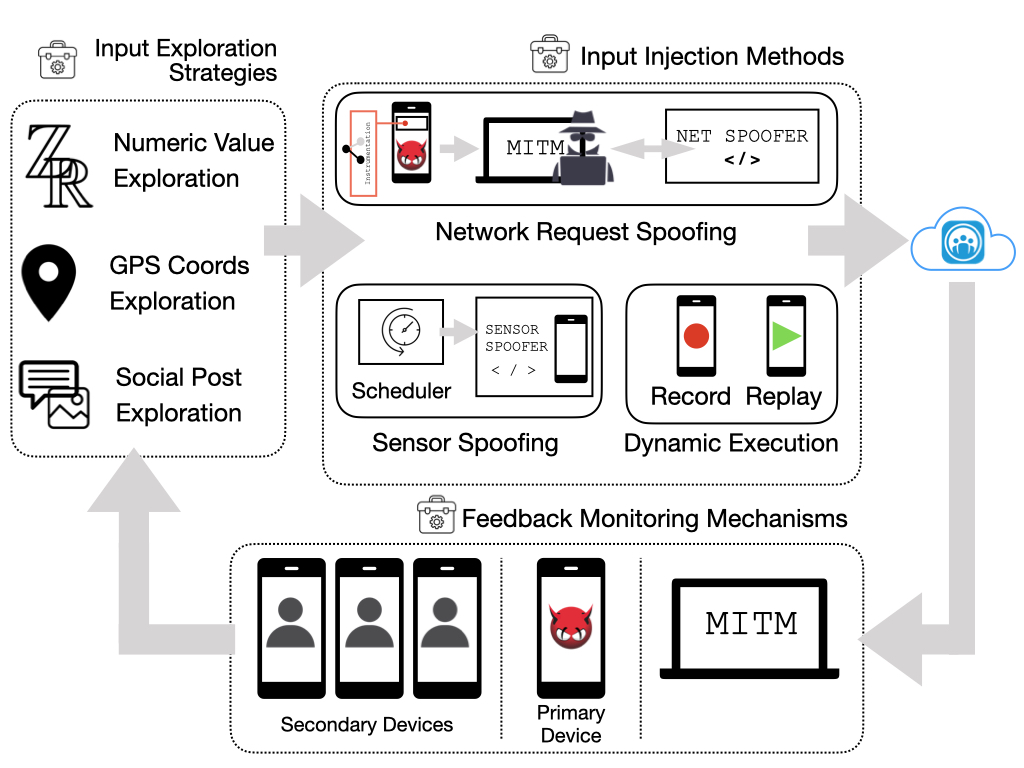}}
    \caption{IIV Analysis Framework.}
    \label{fig:framework}
\vspace{-10pt}
\end{figure}

To overcome these challenges, we present an analysis framework which can aid with the characterization of the IIV attack surface of MCSs (Figure~\ref{fig:framework}). While it still requires some manual effort such as identifying the right APIs (for spoofing network requests) or the right UI elements (for dynamic execution method), our framework can significantly speed up the analysis process compared to a completely manual analysis because it supports range and constraint, and semantic \textit{input exploration strategies}, which drive a set of \textit{input injection methods}. These methods are designed to quickly identify the attack surface of a service by fast repeated injections to the service compared to a completely manual attacker who has to navigate the whole application UI by hand for each injected value. Moreover, since we only have blackbox access to the services, we need a way to verify the success of an injection attempt. To address this, we introduce in our framework \textit{feedback monitoring mechanisms}. Next we elaborate on each of the framework's main components.

\subsection{Input Exploration Strategies.}
\label{subsec:input_exploration_strategies}
Our framework can leverage a wide range of input exploration strategies. These provide the inputs to be used by the injection methods. These strategies are also dynamically informed through feedback monitoring.  For our analyses we devised and implemented three types of strategies supporting range and constraint, and semantic analysis: \textit{numeric value exploration} strategies, a \textit{GPS coordinates exploration} strategies and a \textit{social post generation strategies}.

\vspace{3pt}\noindent\textbf{Numeric Value Exploration (NVE) Strategies.}
Exploring numeric values, (these can correspond to distance, speed, prices, etc.) is tedious when performed manually, and in some cases temporally prohibitive. For example, the range of 32-bit integer values is $-(2^{31}-1) \leq x \leq 2^{31} - 1$. A brute force approach is also impractical. Firstly, most services, to deal with denial of service attacks and to reduce the load on the server-side, rate limit the requests made by clients. For example \textit{Strava} (see Section~\ref{subsec:fitness_strava}) only allows 50 \texttt{POST} requests per account per day. Some of them (see Section~\ref{sec:prices}) even blacklist offending clients. Secondly, some experiments (see Section~\ref{sec:safety}) require dynamic execution and interaction with the target apk. A single injection experiment on \textit{Transit} (see Section~\ref{sec:transit}) takes 15 minutes. Thus, performing all the trials is not efficient. To address this we devise a simple yet efficient strategy to explore the domain range of numeric (integer and float) value injections. 

NVE begins with a geometric growth approach (doubling) starting from an initial value aiming to identify the first value that results in attack failure. It then switches to a linear strategy starting from the last successful injection until  another unsuccessful injection is encountered. More formally, during geometric growth, each value is calculated as $x_i = x_{i-1}\times2$, where $x_i$ is the value to be tried at time interval $i$. Then, during the linear growth, each value is calculated as $x_i = x_{i-1} \pm s$, where $s$ is the step size and ($\pm$) dictates the direction of exploration. The value of $s$ is set to the minimum positive value for integers (i.e. 1) and for floats is set according to semantics. For example, for prices, it is set to represent 1 cent or (0.01). This approach has clear benefits. For example, in a scenario where the positive integer injection success boundary is at 140, linear exploration requires 140 injections. NVE's hybrid approach requires 20 (an 85.7\% percentage reduction): 8 steps to find the first failure value $256=2^8$; and another 12 linear steps ($140-(2^{8-1})$).

\vspace{3pt}\noindent\textbf{GPS Coordinates Exploration (CE) Strategies.}
Geo-sensitive services crowdsource GPS coordinates for a point or event of interest (PoI). GPS coordinates can be derived from a pair of angular measurements, known as longitude and latitude. Longitude represents the east-west geo-position of a PoI on the surface of Earth. Latitude represents the north-south geo-position of a PoI on the surface of Earth. Both angular measurements can be measured in \texttt{degrees}; longitudes range from $-180^{\circ}$ to $+180^{\circ}$; latitudes range from $-90^{\circ}$ to $+90^{\circ}$. CE encompasses four approaches to explore the geo-location ranges for a reported PoI. 

\vspace{3pt}\noindent$\bullet$\textit{ CE-O: Out of range.} CE-O uses \textit{numeric value exploration} to explore longitude and latitude values outside of their expected range. It uses each of the 4 boundaries of longitude and latitude as initial values and explores in the opposite direction of the expected range. CE-O is configured with $s=1$ and explores degrees as integers.

\vspace{3pt}\noindent$\bullet$\textit{ CE-Long: Longitude exploration.}  CE-Long fixes the latitude and explores the longitude range ($-180^{\circ} \leq x \leq +180^{\circ}$), increasing linearly in the positive and negative integer direction.

\vspace{3pt}\noindent$\bullet$\textit{ CE-Lat: Latitude exploration.} CE-Lat fixes the longitude and explores the latitude range ($-90^{\circ} \leq x \leq +90^{\circ}$), increasing linearly in the positive and negative integer direction .

\vspace{3pt}\noindent$\bullet$\textit{ CE-2D: Lat/Long Range exploration.} CE-2D also uses linear exploration but this time to create PoIs over the whole 2D range of longitudes (-180 to 180) and latitudes (-90 to 90). CE-2D uses a step of size $s=5$. Note that 2D does not need to use the hybrid geometric-linear NVE approach since it is temporally feasible to explore the entire 2D range using fixed increments.

\vspace{3pt}\noindent$\bullet$\textit{ CE-Prec: Precision exploration.} This is used to explore the adversary's capabilities at the precision of the fractional part of the longitude/latitude float values. First, we trigger search APIs on the target services to acquire existing PoIs and from those we identify the maximum number of decimal places the service returns for reporting the location of its PoIs. To explore how close two PoIs can be introduced, CE-Prec chooses a starting point (e.g. $(1.0,1.0)$) and explores all combinations of values ($0$--$9$) for the rightmost decimal place for longitude and latitude for a total of 100 injections. If a failure follows a successful injection (too close to the previous injection), we reduce the number of decimal points and repeat.

\vspace{3pt}\noindent\textbf{Social Post Generation Strategies.}
Some services allow their participants to share semantically meaningful information (see Section~\ref{sec:safety}). This information is communicated in the form of posts which can support both text and images. To better understand the extend of the effectiveness of the semantic validation employed by such services we devise three fake post generation strategies. These are designed to explore a target service's level of semantic validation: no validation; general natural language understanding; natural understanding relevant to the service.  

\vspace{3pt}\noindent$\bullet$\textit{ Random Sentence Generation (RSG).} The first strategy aims to explore whether the service accepts any text input without any semantic validation. To verify that, we generate sentences made of words randomly sampled from the English dictionary. The sentence length can be determined by computing the average number of words in genuine posts collected from the target service. 

\vspace{3pt}\noindent$\bullet$\textit{ Sentence Generation with Pre-trained GPT-2 (SGP).}
The second strategy tests posts of semantically similar topics as the extracted posts and a category label under which they are published or organised on the service. To do so, SGP leverages the pre-trained model GPT-2, a state-of-the-art transformer-based model~\cite{radford2019language} to generate the text. The model takes keywords as prompts that it will then complete given its language model. The keywords are selected as the most common words in a set of genuine posts for each category. Those  keywords also become the title of the posts. For the ones that get rejected by the service, SGP uses them again but this time it augments the text with an image. In the first attempt, SGP uses an image irrelevant to all categories. It then uses the rejected posts a second time but this time it augments the generated text with a semantically relevant image. To select relevant images, SGP extracts the most significant entity from our fake text using Google Cloud Natural Language API~\cite{CloudNat6:online} and searches for an image with the entity as a keyword using ``Google Images Download''~\cite{GoogleIm11:online}. It then selects the first search result as the image to use in the post. Lastly, SGP uses text+image-accepted posts again but this time augmented with an image that is irrelevant to all categories. 

\vspace{3pt}\noindent$\bullet$\textit{ Sentence Generation with Adapted GPT-2 (SGA).}
 The third strategy tests more relevant posts generated by fine-tuning the model~\cite{minimaxi63:online} on the set of collected genuine posts. The prompts given to the model now become the category of posts that are to be generated. Similarly with the SGP approach, SGA repeats the generated posts with and without the irrelevant and relevant image. 
 

\vspace{-5pt}
\subsection{Input Injection Methods}
\label{subsec:input_injection_methods}

MCS companion apps, collect data through \textit{sensors} and from their users inputting data in the apps' \textit{UI}. This data is then communicated to the remote services through \textit{Internet-exposed APIs}. We implement three methods  each targeting in aiding the IIV analysis through each of those interfaces:  the \textit{Sensor Spoofing Method} facilitates sensor input analysis; the \textit{Dynamic Execution Method} facilitates UI input analysis; and the \textit{Network Request Spoofing Method}  which facilitates the analysis of inputs directly through the network. 

\ignore{
\begin{figure}[!ht]
    \vspace{-10pt}
    \centerline{\includegraphics[clip, trim=0.5cm 4cm 0.3cm 1cm, width=0.8\columnwidth]{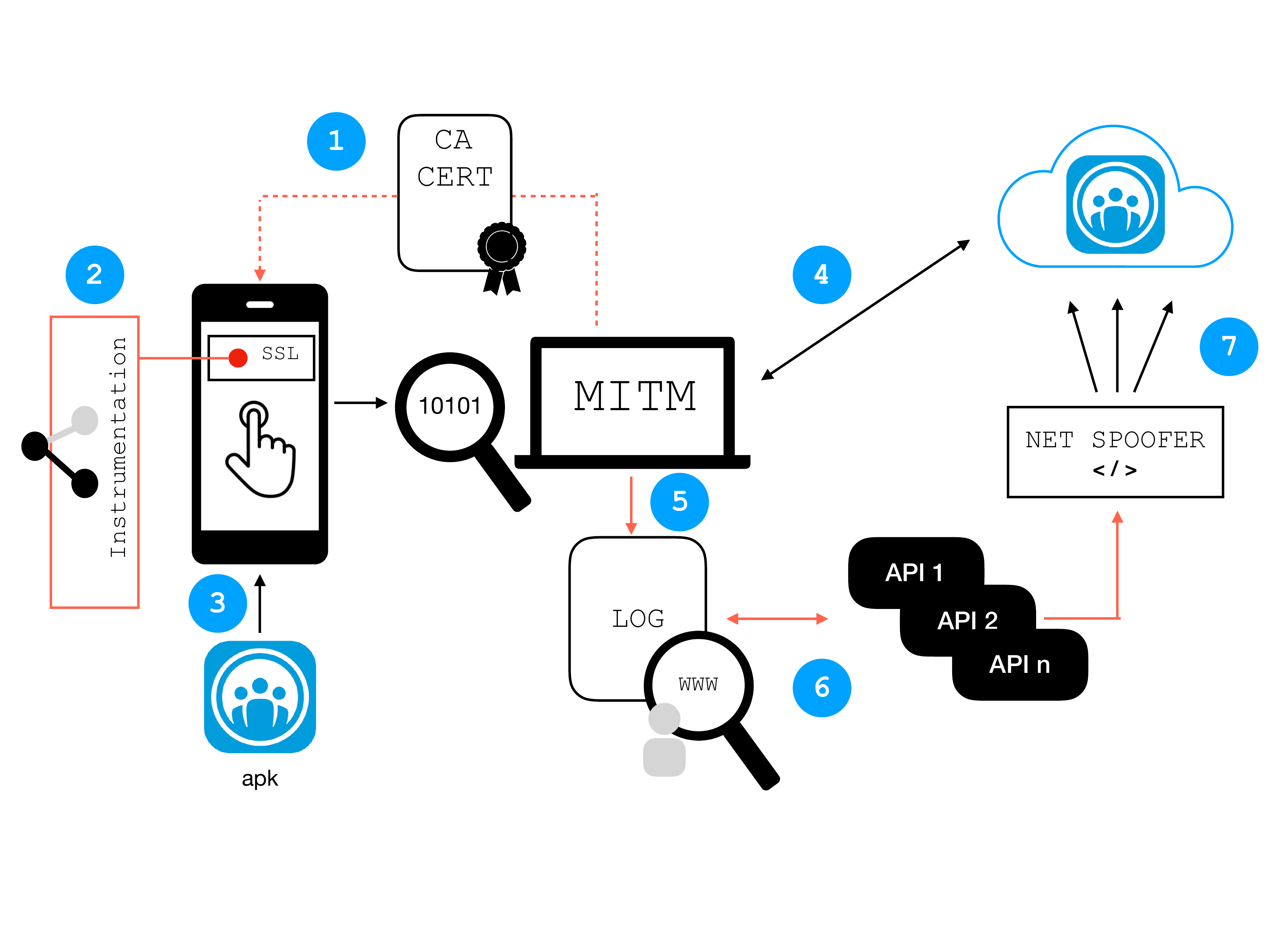}}
    \caption{Spoofing Network Requests workflow.}
    \label{fig:mitm}
    \vspace{-8pt}
\end{figure}
}

\vspace{3pt}\noindent\textbf{Spoofing Network Requests.}
For some apps, we need to analyze their network traffic to extract target requests that can be spoofed for injecting data to the MCS or to monitor its response. In all cases, the communication between the app and the service is encrypted, hence common tools such as Wireshark will not help. To address this we use a man-in-the-middle (MITM) proxy and install its CA certificate on a mobile device. Normally this would have had been enough to enable observing the target apps traffic in plaintext but some apps are using mitigation strategies against MITM attempts and in particular a technique called certificate pinning. This configures the client app to accept connections only with the legitimate server. Thus the app under analysis would reject the proxy certificate. To mitigate this, we use dynamic instrumentation to target and overload the SSL context initialization function (\texttt{SSLContext.init}) of the target app at runtime so that it uses the proxy's certificate and effectively bypassing the app's certificate pinning. Our current implementation uses an extended version of the Android Frida framework~\cite{Frida} to facilitate the dynamic instrumentation. With this setup, we can now run the target app\ignore{\circled{3}} and monitor the network requests it makes and the service's responses. All networking information is logged. This allowed us to reverse engineer the network-exposed APIs of the target services, which are filtered to select the ones to be targeted for injection and response monitoring. Lastly, we developed a network request spoofer, which spoofs network requests to the given APIs, emulating the mobile device, target app, and a user of the service. 




\vspace{3pt}\noindent\textbf{Dynamic Execution Method.} In some cases, injection experiments require dynamic execution of the apps. Performing this manually for a large number of input trials might be prohibitively cumbersome. To address this we use a dynamic execution (DEM) method powered by record and replay tools to facilitate UI navigation and interaction. In particular, we employ tools that can monitor unique IDs of Android app UI elements the user interacts with during recording. When an app's layout is dynamically rendered, the DEM recorder parses the hierarchy tree of the layout and with the help of the analyst identifies the IDs corresponding to the UI elements of interest. Note that it is not mandatory for UI elements to have IDs or even if they do they are not necessarily unique. To address this, the DEM recorder also records auxiliary information related to the UI element of interest such as its position within the hierarchy tree, the element's class, and any textual information (e.g. strings associated with text labels of the element) semantically describing the element. During replay, a DEM replay component leverages the android debug bridge to install and launch the app under analysis on an emulator or real device. It then uses the recorded UI elements' information and their logical order of execution to generate UI interaction commands sent through \texttt{adb} to the app. Like before, if there is a conflict of IDs or absence of them, the DEM replay looks for semantic hints using regular expressions or exact matching on strings, class names, and position within the hierarchy tree to determine which element to emulate an interaction with. Our current implementation of the tools employed in the DEM method use extended versions of the Android UIAutomator~\cite{UIAutoma30:online} during recording and the Appium framework~\cite{AppiumMo3:online} during replay.

\vspace{3pt}\noindent\textbf{Spoofing Sensors.} In other cases, we have to spoof sensor measurements. For example, the transportation service \textit{Transit}, tracks bus rides, during which it continuously reports the GPS coordinates of the host device. Manual attacks~\cite{SIMONWEC66:online} are possible, yet not scalable. To address this, we develop a sensor spoofing module that can be configured to provide fake GPS measurements to a target device. The module uses the Genymotion shell connected to execute commands on a Genymotion Android device emulator~\cite{Genymoti8:online}. We implement a \textit{scheduler} which uses the command \textit{ gps setstatus enabled} to enable GPS readings from the shell. Then it invokes the sensor spoofing module which uses the  \textit{gps setlatitude} and \textit{gps setlongitude} commands to update the GPS coordinates of the emulator according to a time series of GPS values provided by the scheduler. The scheduler generates the time series based on different speeds of movement we want to target (see Section~\ref{sec:transit}).

\vspace{-5pt}
\subsection{Feedback Monitoring.} 
\label{subsec:feedback_monitoring}
Since we only have blackbox access to the services, we need a way to verify the success of an injection attempt. To address this, our framework uses \textit{feedback monitoring mechanisms}. We leverage the fact that most of these services need to provide real-time feedback to their participants through their mobile apps. Thus, on every input injection attempt, we take feedback from the service through (a) secondary spoofed network requests,  (b) through the UI of the primary injection device---adversary's simulated vantage point, or (c) through secondary observer devices registered to the service with a different account---safely simulating victim participants. When feasible, feedback is  observed manually. Otherwise, it is facilitated by our \textit{DEM recorder} and our MITM tools.

\subsection{Analysis Methodology}
To better understand the extend of IIV vulnerabilities we focus our analysis on selected high-profile MCSs spanning various application domains. We regard a MCS to be high-profile if it has a wide userbase and/or it is developed by mature developers/companies. We choose such MCSs since if they exhibit any IIV vulnerabilities these would have the potential to affect a large number of users. At the same time, we expect such MCSs to have a higher responsibility and the resources to deploy security measures.

To identify representative cases, we searched for relevant apps using crowdsourcing related keywords on online search engines. We augment the results by crawling 3259 top free apps from all Google Play categories. This yielded a total of 3295 relevant apps. Two researchers then manually and independently rated each app as either ``mobile crowdsourcing'' (n=112) app or ``other'', resolving conflicts with a discussion. 
Subsequently, the relevant apps were categorized based on their application domain. We identified five domains: Fitness Activity Services, Pricing Services, Transportation Services, Location-Based Services and Safety Services. Then, we cherrypicked at least one representative and previously unexplored high-profile case for our analysis, resulting in a list of 10 apps (Table~\ref{tab:num_injections}).  For each of the cases, we leverage our framework to characterize their IIV attack surface. Sections \ref{sec:fitness}, \ref{sec:prices}, \ref{sec:transit}, \ref{sec:location} and \ref{sec:safety} present our characterization experiments and results on Fitness Activity Services, Pricing Services, Transportation Services, Location-Based Services and Safety Services respectively.  

\begin{table}[htbp!]
\centering
\caption{Number of successful input injections.}
\label{tab:num_injections}
\resizebox{\linewidth}{!}{
\begin{tabular}{l|c|c|c|c}
\textbf{Service}              & \textbf{No. Installations} & \textbf{Domain} & \textbf{Input Injections} & \textbf{Interface} \\ \hline
Strava                        & 10,000,000+ & Fitness Service & 708     & Web API       \\
Fitbit                        & 50,000,000+ & Fitness Service & 98      & Web API       \\
Map My Run                    & 10,000,000+ & Fitness Service & 797     & Web API       \\
Basket Savings                & 100,000+ & Pricing Service & 156     & Web API       \\
ToiFi (Toilet Finder)         & 50,000+ & Location-Based Service & 2728    & Web API       \\
Police Detector               & 5,000,000+ & Location-Based Service & 2910    & Web API       \\
Transit                       & 5,000,000+ & Transportation Service & 403     & Sensor        \\
Neighbors By Ring             & 1,000,000+ & Safety Service &  113    & App UI        \\
Google Maps                   & 5,000,000,000+ & Transportation Service &    -     &     App UI, Sensor     \\
Gas Buddy                     & 10,000,000+ & Pricing Service &   -  & Sensor        \\
\end{tabular}%
}
\end{table}

\ignore{
\begin{table}[]
\centering
\caption{High-profile MCSs selected for analysis.}
\label{tab:apps}
\resizebox{\linewidth}{!}{%
\begin{tabular}{l | l | l | l}
   & Title             & \#Installations & Domain                       \\ \hline
1  & Strava            &                 & Fitness Activities           \\
2  & Fitbit            &                 & Fitness Activities           \\
3  & Map My Run        &                 & Fitness Activities           \\
4  & Basket Savings    &                 & Pricing                      \\
5  & Gas Buddy         &                 & Pricing                      \\
6  & Transit           &                 & Transportation \& Navigation \\
7  & Google Maps       &                 & Transportation \& Navigation \\
8  & ToiFi             &                 & Location-Based               \\
9  & Police Detector   &                 & Location-Based               \\
10 & Neighbors by Ring &                 & Safety                      
\end{tabular}
}%
\vspace{-5pt}
\end{table}
}

\section{Fitness Activity Services}
\label{sec:fitness}
Several MCSs collect fitness information from their participants for better health and wellness insights. For example, Strava Labs~\cite{StravaLa36:online} leverages the large user base of Strava, a service that tracks exercise activities. Moreover, the information collected is used to enable social features such as competing with other participants on local and global challenges. Winners are incentivized with small motivating rewards. Here we explore how an adversary can poison the data collected by such popular services, namely, \textit{Strava}, \textit{Fitbit}, and \textit{Map My Run}, and fake a number of activities, including running, swimming, and cycling, with superhuman performance in terms of distance covered and speed achieved, allowing them to win challenges and rewards. All experiments below are launched through \textit{spoofed networked requests}.

\vspace{3pt}\noindent\textbf{Strava Overview.}
\label{subsec:fitness_strava}
\textit{Strava}~\cite{StravaSu80:online} allows its users to report a number of physical activities, such as running, cycling, and swimming. Its Android app is installed by more than 10,000,000 users. For each of the activities supported, users can report the date and time duration of the activity and distance covered among others. Using a fake account, we were able to fake a running activity covering 50,000 km \ignore{(31068.56 miles)} in 3.5 hours which corresponds to a speed of 14,285 km/h \ignore{(8876.73 mph)}. This constitutes a 98.4\% increase on the world record for the fastest aircraft---7,200km/h \ignore{(4473.87 mph)}. To better understand the extent of these attacks on Strava we design a set of systematic experiments.


\vspace{3pt}\noindent\textbf{Experiment Design.}
Using our MITM proxy testbed (see Section~\ref{sec:framework}) we observed that the Strava app uses a \texttt{POST} request\ignore{ (\textit{https://} \textit{m.strava.com/api/v3/activities})} to submit a new activity to the remote service. The request is bundled with the activity data in JSON format which can include the activity date, duration, distance covered, and a description. We run our experiments by spoofing network requests from a fake account ID which is created when we create a fake athlete's profile. In each trial we use the \textit{numeric value exploration strategy} (~\ref{subsec:input_exploration_strategies}) to check the range of successful injection attacks in terms of \textit{distance} values and \textit{duration} values. Both values are integers. We select the initial value in the exploration to be 0. To study the effect of different types of activities in the input validation we repeat the experiment for 3 types of activities: running, cycling, and swimming.

To detect whether an injection is successful, we leverage another network API. We observe that \textit{Strava} responds to a \texttt{GET} request\ignore{on \textit{https://m.strava.com/api/v3/athletes/userID/stats}}, with the stored statistics of a specific athlete. Thus, issuing this request using our fake athlete's account ID, allow us to check whether the previous injected value was accepted and stored in our profile by the service. In doing the experiments we found that \textit{Strava} only accepts 50 posts by an athlete per day, irrespective of the activity posted. After that, it ignores all posts. To overcome this, we spread our experiments across multiple days.



\vspace{3pt}\noindent\textbf{Results.} Firstly we observe that all three exercise types share the same input domain range boundaries namely (0 to 31,622,400 sec. and 0 to 50,000,000 meters)---at least they do not accept negative values. Unequivocally, these boundaries for duration and distance allow for implausible values. The maximum boundary for the duration (31,622,400 s) corresponds to a run activity which ``took'' 8784 hours or 1 year to complete. Considering that the equatorial circumference of the Earth is 40,075 km, a maximum distance of 50,000 km would correspond to running around the Earth 1.25 times.\ignore{Also, combining the maximum allowed duration and distance, this would correspond to a run speed of 5700mph, or equivalent to 10x the speed of a commercial jet aircraft.} In terms of the maximum distance an athlete can accumulate, we found this to have \textit{no input range restrictions} as \textit{Strava} accepts 4,294,967,295 meters, which is equivalent to the maximum value an unsigned 32-bit integer can have ($2^{32}$). 

An adversary can select values from this range that look plausible but still unequivocally perform better than the top athletes to fake activities and finish first on any challenge leaderboard and claim the rewards (which have monetary value sometimes) and fame that come with it. Moreover, health insurance companies increasingly leverage individuals' health and fitness information to decide the risk score and thus the premium to charge potential and existing clients. IIV vulnerabilities can be exploited to artificially inflate physical activities to manipulate those predictions and thus incur costly damages to insurance companies.

\vspace{3pt}\noindent\textbf{Other Fitness Services.} We conducted experiments to study the resilience of other popular fitness services, specifically \textit{Fitbit}~\cite{FitbitOf86:online} and \textit{Run with Map My Run}~\cite{MapMyRun46:online}. Regarding \textit{Fitbit}, we found that activity distance can be added from 1km to $1609.344km$ (equivalent to $10,000$ miles). For activity duration, we observe that there is no input data type positive range restriction as we could inject duration from 1 second to $2,147,483,647$ (which is the maximum positive value for a 32-bit signed binary integer, or $2^{31}-1$). Regarding \textit{Run with Map My Run}, alarmingly, we found that any input validation happens on the client-side while the values are stored non validated (we injected arbitrarily large duration values) on the server. Due to space limitations, we defer further details to Appendix~\ref{app:other_services}. 

\section{Pricing Services}
\label{sec:prices}
Pricing crowdsourcing services, allow their participants to report the exact price value of an asset. Tampering with these prices can lead to an unfair competition where customers are driven away from target stores or directed toward particular stores. In this section we elaborate on a practical attack we launched against a popular pricing crowdsourcing service (\textit{Basket Savings}). The attack is launched through \textit{spoofed network requests}.

\vspace{3pt}\noindent\textbf{Manual Analysis Description and Objectives.} \textit{Basket Savings}~\cite{BasketSm94:online} is an example of a money-saving service which crowdsources prices of grocery items on superstores. Its Android mobile app was downloaded more than 100,000 times. The app allows users to add a price (float value) for an item only if they are within a GPS-determined geo-location close to the target store and by scanning the bar-code of the target product or their purchase receipt. Moreover, the \textit{Basket Savings} service blocks devices by IP in case they detect suspicious activity like a user making malicious price changes on the app. Nonetheless, we found that one can bypass this by leveraging a feature on the app which allows users to manually input and submit prices through the app's user interface. We suspect that this feature was added to increase crowdsourcing opportunities, for example by supporting price reports when the user has left the store. We verified that an adversary can use an emulator device to inject both lower and higher than the real price values for selected target products. For example, we verified that we could increase the price of one gallon of milk by 129\% of its usual price (from \$3.49 to \$8). As with the case of Transit, we monitor feedback to verify the results by accessing the values from secondary passive devices registered as users of the service. This, can be leveraged by an adversary to launch an unfair competition attack where customers are driven away from target stores or directed toward particular stores by manipulating the advertised prices for popular products.

\begin{table}[!t]
\centering
\scriptsize
\caption{Basket: Trader Joe's \& Amazon Prime(*)}
\label{tab:basketcompare}
\resizebox{\columnwidth}{!}{%
\begin{tabular}{l | l | l | l | l | l | l} \toprule
\textbf{Product}                           & \textbf{Value} & \textbf{Min} & \textbf{Max}  & \textbf{*Value} & \textbf{*Min} & \textbf{*Max} \\ \hline
Apples                           & 0.49 & 0.05 & 2.0 & 1.58 & 0.16 & 4.0 \\
Bananas        & 0.19 & 0.09 & 2.0 & 0.55 & 0.06 & 2.0 \\
Strawberries        & 0.99 & 0.09 & 2.0 & 5.0  & 2.21 & 8.3 \\
Eggs                 & 1.99 & 0.2  & 4.0 & 2.12 & 0.21 & 6.0 \\
Chicken Breasts     & 2.69 & 0.27 & 6.0 & 3.25 & 0.33 & 8.0 \\
Organic whole Milk & 3.49 & 0.35 & 8.0 & 3.76 & 0.38 & 8.0 \\
\bottomrule
\end{tabular}
}%
\vspace{-5pt}
\end{table}

\begin{table}[!t]
\centering
\scriptsize
\caption{Basket: Milk on Trader Joe's}
\label{tab:basket_traderjoes_milk}
\resizebox{\columnwidth}{!}{%
\begin{tabular}{l | l | l | l | l} \toprule
\textbf{Product} & \textbf{Gallons}  & \textbf{Value} & \textbf{Min} & \textbf{Max} \\ \hline
Whole Milk 1      & 0.5       & 1.29 & 0.13 & 4.0 \\
Whole Milk 2      & 0.5      & 2.29 & 0.23 & 6.0 \\
Organic Whole Milk 1   & 0.5    & 2.99  & 0.30 & 6.0\\ 
Organic Whole Milk 2      & 1         & 5.69 & 1.71 & 10.58 \\
Homogenized Whole Milk & 1   & 5.99 & 1.80 & 6.59 \\
\bottomrule
\end{tabular}%
}
\vspace{-10pt}
\end{table}

\vspace{3pt}\noindent\textbf{Experiment Design.}
The case above demonstrates that \textit{Basket Savings} is vulnerable to improper input injections. Next, we design a set of experiments to characterize the IIV attack surface of the service. Prices in Basket are reported as float values. To find the input domain range accepted by the service---and thus the range of the adversary---we perform price injection attacks on the user interface of its mobile app, aiming to identify the boundaries (minimum and maximum values) the adversary can inject. Trying all possible values for an input type can be very time-consuming. In this case, a float data type has a range $-3.4E+38 \leq 3.4E+38$. Therefore to find the accepted input range in a more efficient manner, we follow the numeric value exploration strategy outlined in Subsection~\ref{subsec:input_exploration_strategies}. We choose the initial value to be the current value of a target product. We also choose the step size to be equivalent to 1 cent ($s=0.01$).

To perform the injections, we used our MITM-proxy testbed to obtain the API call for submitting a new price value for any given store. We observed that the API call\ignore{was the \textit{POST} request \textit{'https://api.basketsavings.com/pricesnap/app/product/} \textit{price/add'} and it took} uses the new price, product and store-id along with a longitude, latitude pair and a timestamp. We wrote a script that could replay this API call to Basket's server and used it manually for injecting different prices for any product according to the aforementioned numeric exploration approach.

To detect the success or failure of an injection trial, we used another API call discovered through our MITM-proxy testbed\ignore{ ( \textit{GET} request \textit{https://api.basketsavings.com/basket-api/product/analysis?} \textit{productId=207862\&radiusMiles=25})}. Using a secondary passive device we manually verify the injected value was visible for other participants and has replaced the prior price for the given item on a particular store. We used this injection and observation approach to verify the \textit{minimum and maximum value} possible for ``bananas'' and ``strawberries'' at the two stores. Later, we discovered another API call that had these maximum and minimum allowed price values i.e. the acceptable price range embedded in its response\ignore{ (\textit{GET} request \textit{'https://api.basketsavings.com/basket-api/oob?productId=208017\&} \textit{storeId=179077'})}. We used this request to verify and obtain the price ranges quoted in the results for this app.

To examine the effect of \textit{location}, we repeat this experiment for two stores (\textit{Amazon Pantry} and a \textit{Trader Joe's}\ignore{at 3131 S Hoover St Ste 1920,} store in Los Angeles, CA). To examine the effect of \textit{product type}, for each store we try manipulating prices for 6 different products: apples, bananas strawberries, eggs, chicken breast, and organic while milk. Lastly, to examine the variation within product types we select 5 different kinds of organic milk.


\vspace{3pt}\noindent\textbf{Results.}
Table~\ref{tab:basketcompare} summarizes the range of successfully injected prices for Trader Joe's and Amazon Prime. We observe that the minimum value allowed for both stores is mostly the 10\% of the current value and the maximum value allowed for any product mostly seems to be $2 * \lceil current Value \rceil)$. These boundaries are clearly not realistic. Note that only bananas on Trader Joe's and strawberries on Amazon show exception to these rules in the results. Table~\ref{tab:basket_traderjoes_milk} shows the successfully injected prices for 5 different kinds of organic milk on Trader Joe's (these are different from the one listed on Table~\ref{tab:basketcompare}). We observe that the minimum allowed price was 10\% of the shown price for milk under \$5 and roughly 30\% of the shown value for milk over \$5. The same rules as above were followed for 3 out of 5 kinds of milk for the maximum price.  

\ignore{
\begin{table}[]
\centering
\caption{Basket: Trader Joe's}
\label{tab:basketTraderjoes}
\begin{tabular}{l | l | l | l} \toprule
\textbf{Product}                           & \textbf{Value} & \textbf{Min} & \textbf{Max} \\ \hline
Apples                           & 0.49 & 0.05 & 2.0 \\
Bananas        & 0.19 & 0.09 & 2.0 \\
Strawberries        & 0.99 & 0.09 & 2.0 \\
Eggs                 & 1.99 & 0.2  & 4.0 \\
Chicken Breasts     & 2.69 & 0.27 & 6.0 \\
Organic whole Milk & 3.49 & 0.35 & 8.0 \\
\bottomrule
\end{tabular}
\end{table}

\begin{table}[]
\centering
\caption{Basket: Amazon Prime}
\label{tab:basketAmazon}
\begin{tabular}{l | l | l | l }  \toprule
\textbf{Product}                           & \textbf{Value} & \textbf{Min} & \textbf{Max} \\ \hline
Apples             & 1.58 & 0.16 & 4.0 \\
Bananas            & 0.55 & 0.06 & 2.0 \\
Strawberries       & 5.0  & 2.21 & 8.3 \\
Eggs               & 2.12 & 0.21 & 6.0 \\
Chicken Breasts    & 3.25 & 0.33 & 8.0 \\
Organic whole Milk & 3.76 & 0.38 & 8.0 \\
\bottomrule
\end{tabular}
\end{table}
}

\vspace{3pt}\noindent\textbf{Other pricing services.} For our analysis we also selected \textit{GasBuddy}. \textit{GasBuddy} crowdsources real-time gas stations' fuel prices through their app. Like \textit{GoogleMaps}, it uses client certificate pinning so we couldn't decrypt and reverse engineer the API calls used by the app. Another challenge was the variable nature of \textit{GasBuddy}'s user interface where random pop-up screens (e.g. ads) would cause the app to crash or frequently end up at unintended screens, which currently our DEM module cannot handle.
\section{Transportation Services}
\label{sec:transit}
\textit{Transit}~\cite{Transit:online} is a public transportation service for 175 cities. Its Android app enjoys over 5,000,000 installations. By crowdsourcing every passenger’s real-time \textit{sensory data} including positioning information and speed of movement, Transit can help its users plan a trip and support them in their travel by predicting the expected arrival time of the next subway or bus. However, we found that it is possible to fool the Transit service to accept fake measurements. To demonstrate this, we performed an experiment using 3 Android device emulators. The first device ($E_{\alpha}$) acted as the adversarial device aiming to fool the service, while we used the other two as observer devices ($E_{o1}$ and $E_{o2}$) for verifying the result of the attack on the service on other users' devices. We installed Transit on all three emulators and used the app to plan a bus trip from point A to point B in London, UK. $E_{o1}$ and $E_{o2}$ started their trip 3 bus stops later than $E_{\alpha}$. $E_{\alpha}$ was driven by our sensor spoofer, which was automatically feeding fake GPS updates to the device, simulating a movement along the target bus route with a steady speed (12km/h). 
In this manner, Transit was instantly fooled to accept that $E_{\alpha}$ is actively riding the target bus, which we could verify as the Transit app rendered a new bus icon moving at $E_{\alpha}$ speed and direction, on the screen of all three devices. We were also successful in manipulating the expected bus arrival time for $E_{o1}$ and $E_{o2}$, by faking the speed of movement of $E_{\alpha}$ to be 575mph which is equivalent to the average speed of a commercial plane. All experiments were performed at an off-peak time in a rural area to avoid affecting real users. Competing transit agencies can use these attacks to deter customers from using another transit service. 

\vspace{3pt}\noindent\textbf{Experiment Design and Results.} Next we design a set of experiments to better understand the range and predictability of values an adversary can leverage for manipulating bus routes. We leverage our DEM method (see Subsection~\ref{subsec:input_injection_methods}) to dynamically install and execute the Transit app on Genymotion non-root emulators and move to a target screen for enabling active route navigation in the app. Then we use our sensor spoofing method to fake sensor GPS values to the victim app. All our experiments are performed targeting the same bus route in a rural area. We configure the attacker's device starting location to be 18km earlier on the bus route than the victim device's location (also on the bus route). The scheduler is configured to emulate the \textit{speed of movement} of the adversarial device by generating GPS timeseries values corresponding to different frequencies. An injection attempt at a specific speed of movement is considered successful when it can affect the expectation of the bus arrival time on the observer device. 

\begin{figure}[!t]
\centering
\subfloat[Text element hint on adversarial device's UI.]{%
  \includegraphics[clip, trim=0cm 17cm 0cm 0cm, width=0.7\columnwidth]{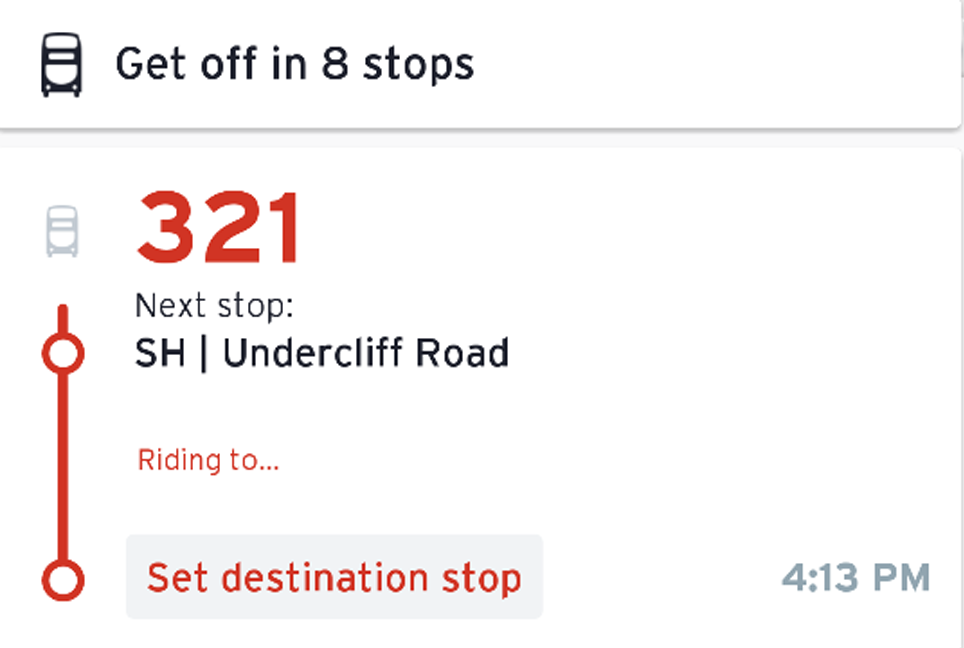}
}\\
\subfloat[Icon hint on victim device's UI.]{%
  \includegraphics[clip, trim=3cm 9cm 3cm 6cm, width=0.7\columnwidth]{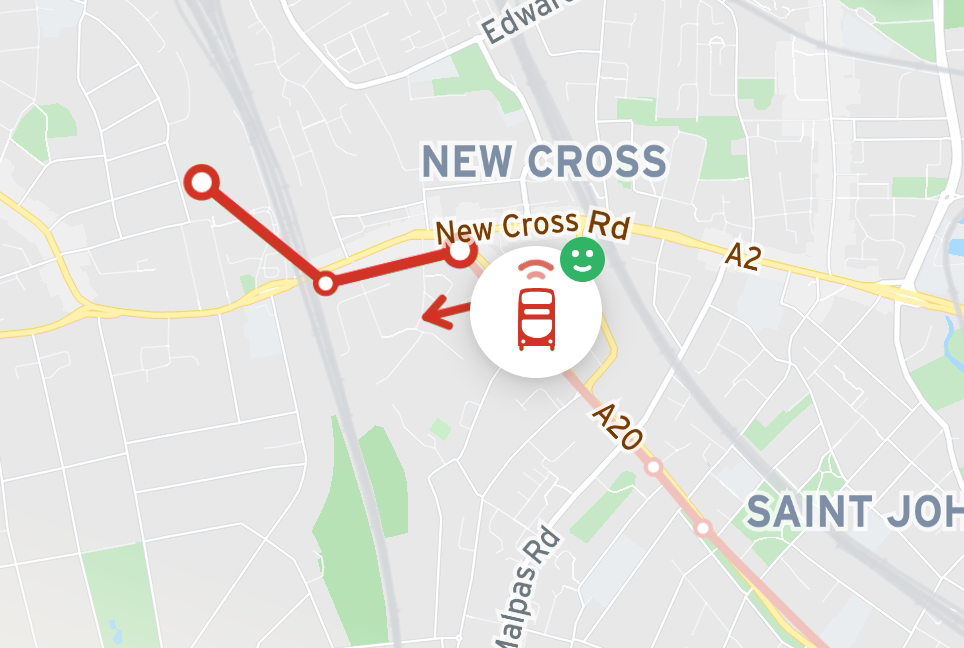}%
}
\caption{Transit mobile app UI hints.}
\label{fig:transitUI}
\vspace{-10pt}
\end{figure}

\vspace{3pt}\noindent$\bullet$\textit{ Linear value exploration.} We first use linear numeric exploration (see Subsection~\ref{subsec:input_exploration_strategies}) to generate speed values from 0 to 1000 km/h with a step size of 10 km/h ($s=10$). We found that 97/100 (97\%) fake movements succeeded in fooling \textit{Transit} that the adversarial device is actively riding a fake bus. We determine success by tracking visual hints on the UI of the app on the adversary's device: \textit{Transit} shows a textual description to its user when waiting for a bus (``Waiting for the bus''). When the user is riding a bus this textual hint changes to ``Get off in [number of] stops'' as shown in Figure~\ref{fig:transitUI}(a). Using our DEM module (see Section~\ref{sec:threats}) we can track the target UI element with the textual hint to verify the success or failure of the experiment. However, we observe that even when the adversary succeeds to create a fake bus, this is not always reflected on other users' devices, especially when moving at high speeds. We verify this by randomly selecting 20 values of speeds and repeat the experiments for each of them, this time also monitoring the observer device. We can confirm that the fake bus also appears on the observer devices by looking for the bus icon with a happy rider face (see Figure~\ref{fig:transitUI}(b)). The exact icon can be extracted beforehand as it is stored in the victim app's \texttt{res/drawable} folders which we access by decompiling the app using an Android reverse engineering tool (apktool~\cite{ApktoolA92:online}). This then used during the experiments with the \texttt{MatchTemplate()} function of the \textit{openCV2} library~\cite{OpenCV93:online} to search for that icon within a grayscale version of screenshots of the UI of the victim's device. Using this approach, we verify that 17/20 (85\%) successful bus fakes also appear on the victim device.

\begin{figure}[!t]
    \centerline{\includegraphics[clip, trim=0.0cm 0.0cm 0cm 1.9cm, width=0.9\columnwidth]{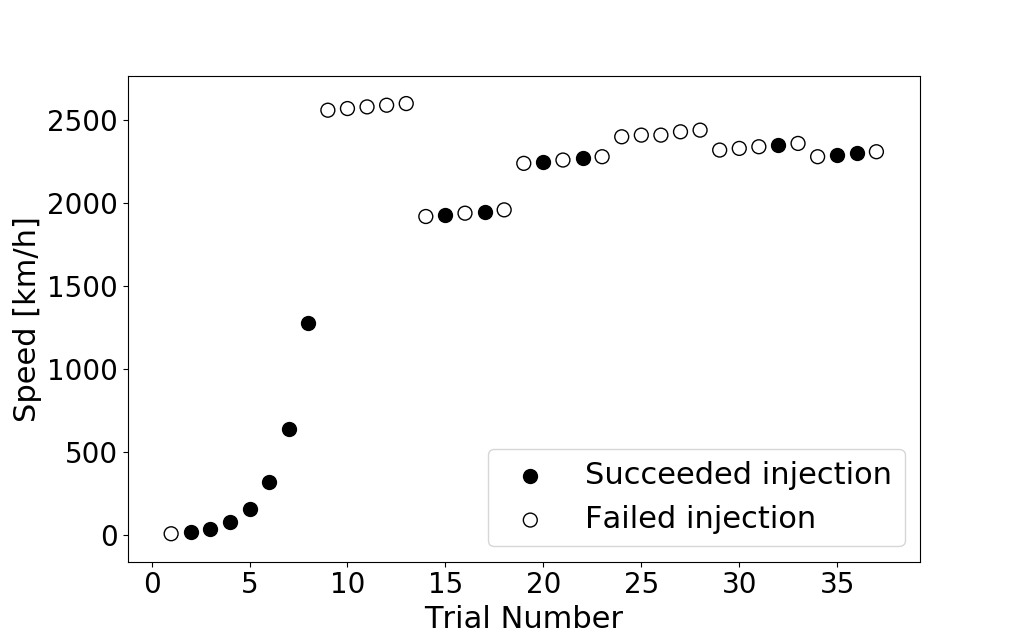}}
    \caption{Transit: Faking Buses with Supersonic Speeds.}
    \label{fig:transitMax}
\vspace{-10pt}
\end{figure}

\vspace{-0pt}\noindent$\bullet$\textit{ Supersonic Speeds.} In this experiment, we aim to find whether supersonic speeds (greater than the speed of sound---1235km/h) are possible. To explore this we use a variation of the numeric value exploration strategy. However, when a failure is encountered during geometric growth, before reverting back to the last successful value, we linearly ($s=10$) try the next 4 values, and only if all 4 fail we finish exploring higher values. This is needed to deal with uncertainties at high speeds where we found the behavior of the service can be unpredictable. Specifically, our algorithm proceeds as follows: to find the highest acceptable speed $S_h$, the attack emulator starts at the speed 10 km/h and keeps doubling this value until the first failed attack speed $S_i$ is observed on the victim phone. The failed attack is further confirmed with 4 more adversarial speeds : $S_i + 10$, $S_i + 20$, $S_i + 30$, $S_i + 40$. If more than 4 attacks fail out of 5 experiments, we regard $S_i$ as the first failed attack speed and $S_{i-1}$ as the last successful attack speed. Therefore, $S_h$ is within the range $[S_{i-1}, S_i]$. Given $S_j = \frac{S_{i-1} + S_i}{2}$, range [$S_{i-1},S_i$] is divided into two ranges: [$S_{i-1},S_j$] and [$S_j,S_i$].  If $S_j$ is able to attack the victim phone, $S_h$ is within  $[S_{j},S_{i}]$, otherwise, [$S_{i-1},S_j$]. By keeping on dividing the range, $S_h$ is finally confirmed. As shown in Figure~\ref{fig:transitMax}, we managed to succeed with supersonic speeds of up to 2350 km/h. We repeated the attack at this speed 10 times. We found that 3/10 (30\%) of the times the injection at 2350 km/h influences the victim.

\vspace{3pt}\noindent\textbf{Other transportation services.} For our analysis we also selected \textit{GoogleMaps} (Section~\ref{sec:framework}). Even though its susceptibility to general data poisoning attacks was established manually~\cite{SIMONWEC66:online}, we could not apply our framework to scale up the analysis. In particular, we could not decrypt its APIs calls due to the usage of not only server but also client certificates which our framework cannot currently bypass. However, our framework can be expanded with a farm of real phones and emulators to support the analysis of such cases.


\section{Location-based Services}
\label{sec:location}
Services such as \textit{ToiFi (Toilet Finder)} and \textit{Police Detector} , crowdsource points of interests (PoIs). An adversary targeting such services might create or remove PoIs to their own benefit. For example, a fake toilet might be used to lure potential victims to deserted locales; a fake police radar can deceive an individual into using another route. Using \textit{spoofed networked requests} we were able to verify and analyze IIV vulnerabilities for both \textit{ToiFi} and \textit{Police Detector}. 

\vspace{3pt}\noindent\textbf{ToiFi (Toilet Finder) Overview}
\textit{ToiFi}~\cite{PublicTo19:online} is a location-based MCS which allows users to find public restrooms  (as PoIs) in time of need. It also has an option for the community to help by adding new PoIs or editing/removing previous PoIs. Its Android app was installed more than 50,000 times. We were successful in both manually adding and removing PoIs at arbitrary locations of interest. To systematically characterize the attack surface of the service next we use the \textit{GPS coordinates exploration} strategy to discover the range of coordinates that can be used to fake a public toilet.

\vspace{3pt}\noindent\textbf{ToiFi (Toilet Finder) Experiment Design and Results.} To systematically characterize its attack surface we use the \textit{GPS coordinates exploration} strategy to discover the range of coordinates that can be used to fake a public toilet. To conduct the experiments we identified---through our MITM proxy---the API the remote service exposes for receiving a request to add a PoI. We then carefully craft network requests triggering the API, by spoofing the mobile app and a service user. Success of the injection is determined by spoofing a second request targeting a different exposed API which allowed us to search for the presence of a PoI. We found that out of bound longitudes and latitude values (CE-O) are rejected. However, all trials on \textit{CE-Long}, \textit{CE-Lat} and \textit{CE-2D} were successful allowing us to inject PoIs anywhere on the surface of the Earth. For \textit{CE-Prec}, we saw that we can inject POIs up to a precision of 7 decimal places for both longitude and latitude. Furthermore, we noticed that there is no check preventing two points from being too close on the map as the first round of injections went fine as it is. Evidently, this can be exploited for fun, for example by injecting toilets in the middle of the desert or in the middle of the ocean. It can also be used for pranks as public toilets can be injected on a target individual's or enterprise's private premises; or even for harm, as people can be lured in dark alleys and isolated areas. We marked all PoIs added by our experiments as non existent by spoofing another API call we discovered so as to not harm the service or its users.

\vspace{3pt}\noindent\textbf{Police Detector Overview.}
\textit{Police detector (Speed Camera Radar)} \cite{PoliceDe94:online} ia another location-based MCS which uses crowdsourcing to help users make intelligent decisions while driving. Its Android app was installed more than 5,000,000 times. Its users can report the location of police speed detectors, road repairs and road accidents. As with \textit{ToiFi (Toilet Finder)} we were successful in faking all PoIs:  fake police speed radars, fake road accidents and fake road repairs. As before, to better understand the susceptibility of the service to these attacks, we perform experiments using the \textit{GPS coordinates exploration} strategies (see Subsection~\ref{subsec:input_exploration_strategies}). 

\ignore{
\begin{figure}[htp]
\centering
\subfloat[using E2 and E3 strategies.]{%
  \includegraphics[clip, trim=0cm 3cm 0cm 3cm, width=0.8\columnwidth]{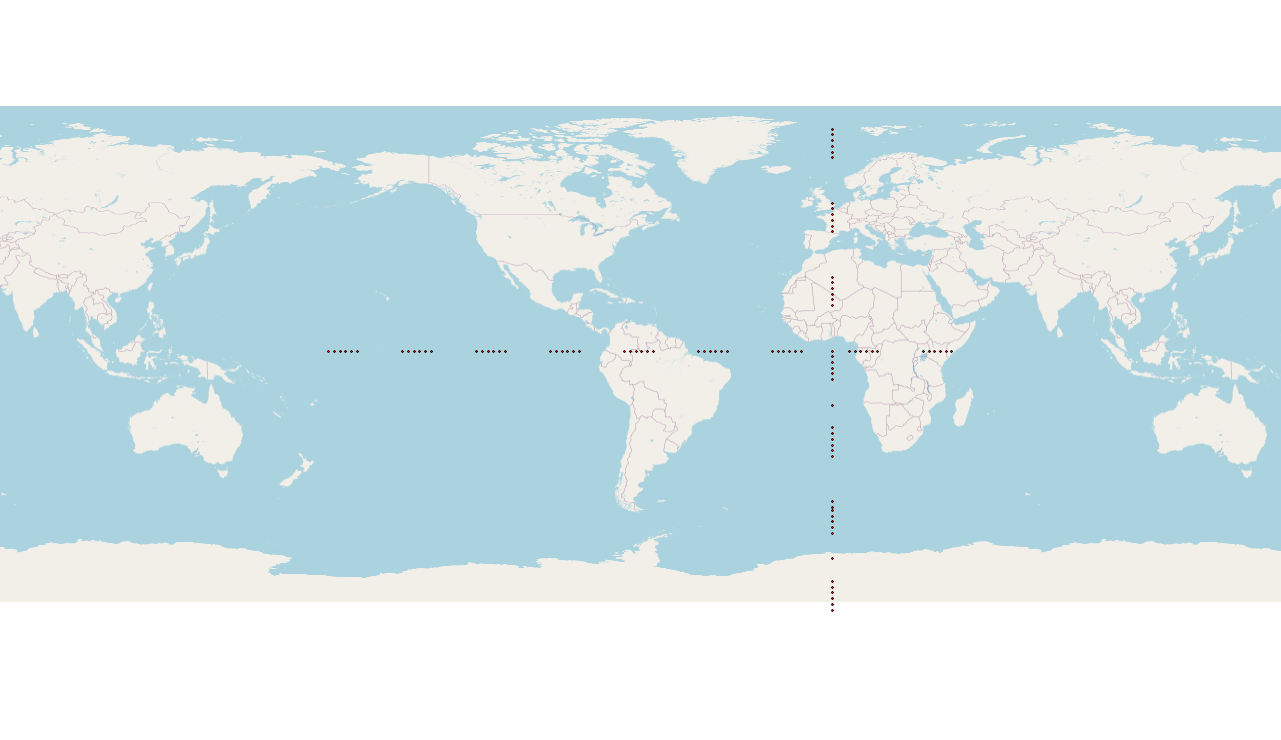}
}\\
\subfloat[using E4 strategy with fake user rotation.]{%
  \includegraphics[clip, trim=0cm 3cm 0cm 3cm, width=0.8\columnwidth]{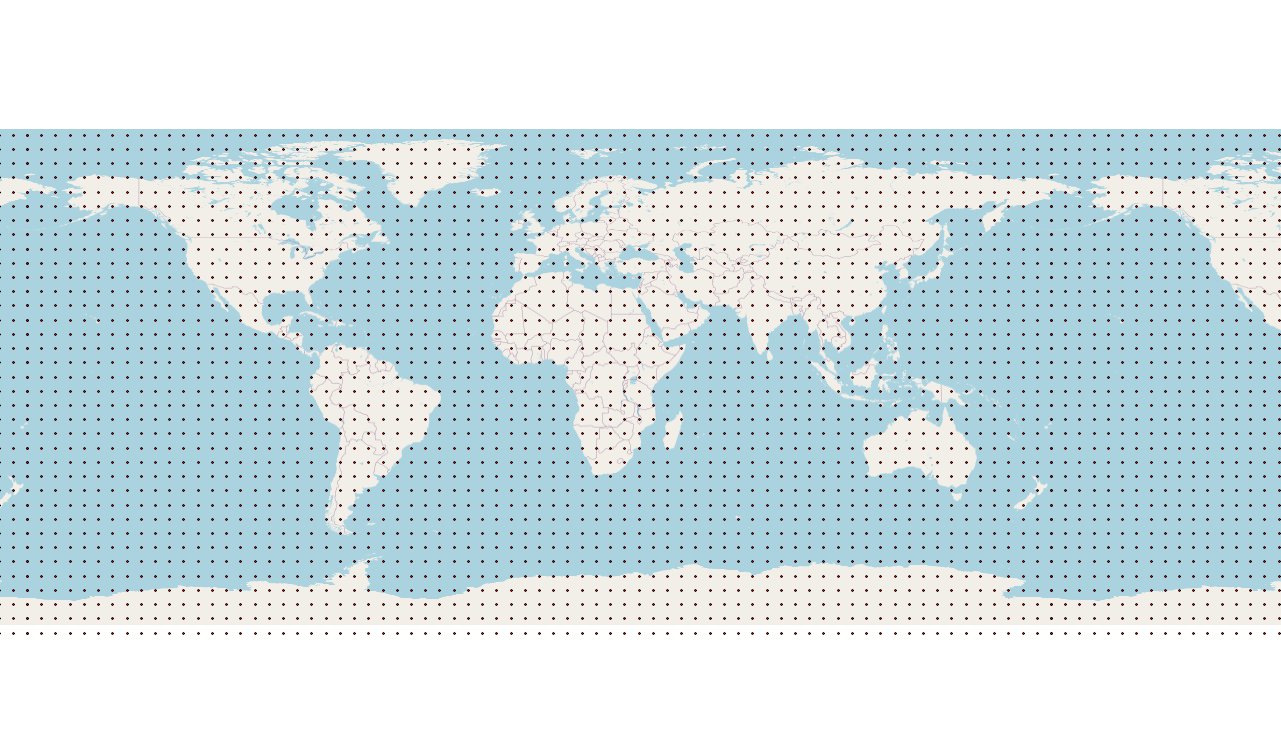}%
}
\label{fig:e234}
\caption{Police Detector: Successful Injections ($\bullet$).}
\vspace{-5pt}
\end{figure}
}

\vspace{3pt}\noindent\textbf{Police Detector Experiment Design and Results.} Using our MITM proxy setup, we reverse engineer the API of the service and identified an API call used to add a speed detector and a second API which allows searching for speed detectors. The second is useful for observing the success or failure of the injection. We successfully spoofed the service's mobile app's network requests using our framework. We found that values outside the expected longitude and latitude range (\textit{CE-O}) are rejected. However, for the latitude and longitude experiments (\textit{CE-Long} and \textit{CE-Lat}), 48/180 (26.6\%) and 55/360 (15.3\%) of the injections succeeded respectively. Consecutive injections are blocked after a fixed number of requests (see red points on Figure~\ref{fig:e234}), which led us to hypothesize that the points were not rejected based on semantic validation but instead were rejected due to a rate limit on the number of points a registered user can submit. To overcome this, we generate a pool of fake user fingerprints which we rotate through when performing the injections. We noticed that the user-id for a participant depends on the Android-id of her phone and relaunching the app after changing the Android-id of the emulator phone creates two requests with two different but corresponding ids for registration and retrieval of PoIs on the map. So, we wrote two scripts to automate this process, one of them employed \texttt{adb} to keep relaunching the app and the other script interacted with our MITM proxy to scan the requests and responses and extract the (registration-id, user-id) pairs out of them so that we can use them later for our experiments. We follow this setup to generate a pool of 86 fake users and conduct the \textit{CE-2D} experiment where we do injections for the whole 2D range of latitudes and longitudes. Furthermore, we add a delay of 10 seconds between successive injections and repeat the whole sequence of requests that happen on launching the app for each injection, instead of just calling the injection API call. With this setup, we were able to perform successful injections for the entire 2D range of points as described in \textit{CE-2D} (see Figure~\ref{fig:e234} (green points)). The only exceptions where the injection failed was when either the longitude or latitude was 0 or when the value of latitude was exactly on the boundaries -90 or 90. For \textit{CE-Prec}, we saw that POIs can be inserted with a precision of up to 5 decimal places but no two POIs can be closer than $0.002$ on either longitude or the latitude scale, which happens to be equivalent to around 222 meters in distance. To prevent any harm to the service or its users we identified another API request\ignore{ (http://142.11.196.19/input.php)} which we employ with an input value of 0 in the request body, to remove the added PoIs after each successful injection. 

\begin{figure}[!t]
\centering
  \includegraphics[clip, trim=0cm 2.5cm 0cm 3cm, width=0.9\columnwidth]{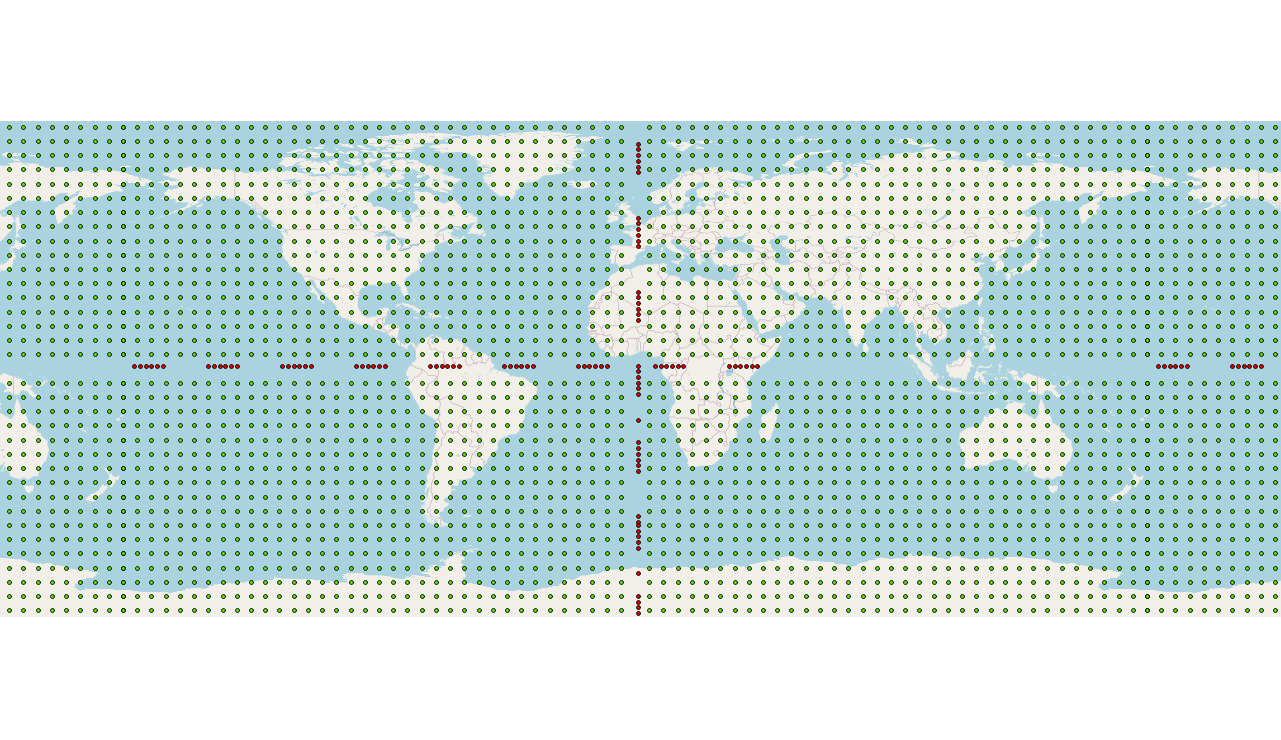}
\caption{Police Detector: \textit{CE-Long} \& \textit{CE-Lat} Successful Injections
(\textcolor{red}{$\bullet$}), \textit{CE-2D} Successful Injections (\textcolor{green}{$\bullet$}).}
\label{fig:e234}
\vspace{-10pt}
\end{figure}

\section{Safety Services}
\label{sec:safety}
A number of MCSs allow users to share safety-related information. One such service is \textit{Neighbors by Ring}~\cite{Neighbor60:online} (NbR), whose Android app has been downloaded over 1,000,000 times. This service allows its users to share four kinds of posts: crimes, safety-related events, lost pets, and unexpected activities. These can include text, images, even video streams from security cameras, and share the reporting device's location. Fake reports can be used to spread chaos or defame a location or neighborhood. In a more sophisticated scenario, enemy states or organizations with political affiliations can deploy elaborate propaganda schemes.

\vspace{3pt}\noindent\textbf{Manual Attack.}
We submitted manually--constructed sample textual posts controlling for both the location and the semantics of the text. We found that NbR does not verify the location but does verify the semantics of the textual report. For example, our attempt to inject ``dangerous cat spotted'' was rejected, but our attempt to inject ``dangerous dog spotted'' was successful. Additionally, posts that were too vague about the safety issue were rejected as well. We suspect that NbR uses a machine learning algorithm to determine whether posts are legitimate or not in order to tackle fake posts.

\vspace{3pt}\noindent\textbf{Experiment Design.} 
To further explore the service's semantic validation, we submit posts generated using our post generation strategies: random sentence generation (RSG), sentence generation with pre-trained GPT-2 (SGP), and sentence generation with adapted GPT-2 (SGA),  and report the posts' acceptance rate. To configure the strategies' parameters, we first use our DEM module to execute the app, interact with it, and extract already present but unseen posts. We collected a dataset of 1080 genuine posts which follow the format: $<$Category: $category_p$, Title: $title_p$, Description: $description_p$ $>$. Using the genuine posts descriptions we determine the average sentence length ($=30$) words for the RSG strategy. We also identify the three most common words present in the first sentence of each post's description, by category. These are used as the keywords and titles of the fake posts in the SGP strategy. For the adaptive approach (SGA), we fine-tuned the text generation model. The model was trained for 1000 epochs and with a learning rate of $10^{-4}$ and a temperature of 0.7 for the generation. 





To test the generated posts, we set up three Android devices and submit them via the UI of the target service's Android app, using our DEM module. Since the app requires a unique email address to create an account, we create temporary emails for each device. We set the devices' location in Death Valley, California at different spots such to minimize exposure to real users. Once a post is submitted, our script updates the page for up to 8 minutes until the post appears in the user's submissions. We noticed that the time to get the decision varies between 1 to 7 minutes. If the post appears, it is marked as accepted and the script deletes the post. If the post is not accepted, the respective email address receives a rejection email and the post never appears in the submissions. We also found that the service blocks accounts that post too often and set a limit of 8 rejected posts a day. Once a user is blocked, their submissions are ignored without any notification. To overcome this problem, we submit at most 3 posts an hour and set random delays in the range of 20 to 35 minutes between posts. We also monitor the emails manually to check that the posts are not ignored.   

\vspace{3pt}\noindent\textbf{Results.}
Table~\ref{tab:accpoststxt} shows the number of accepted posts per category out of the total number of submitted fake posts for each of the generation strategies. Examples of successful injections are shown on Table~\ref{tab:accfakex}. The results show that the app does not accept just any input text, as none of the random posts went through. Furthermore, some text-only posts generated with the \textit{SGP} strategy are rejected even if they are of topics similar to the ones accepted by the service. Those posts were possibly too vague about safety issues and without a clearly defined structure. However, 23 out of 100 fake posts were indeed approved. Fake posts generated using the \textit{SGA} strategy were more likely to get accepted as 66\% of posts were able to replicate the format and or the necessary information required by the app. Moreover, the model generated crime related posts with almost 90\% success rate. The ``Unexpected Activity'' category seems to be the hardest to imitate as it has the lowest acceptance for both strategies. When comparing fake and genuine posts, we found that ``Unexpected Activity'' posts often also include video footage captured with the Ring security camera. 

Enriching posts with \textit{irrelevant} images did not help except for one post that mentions a troop of policemen. We hypothesize that the blue color of the image could be mistakenly taken as their usually blue uniform. Nevertheless, this experiment show that there is semantic validation using the images. Indeed, posting the fake reports with \textit{relevant} images improved the success rate for both the \textit{SGP} and the \textit{SGA} strategies and particularly for the category ``Lost Pet''. The rates also improve slightly for the other categories.


\begin{table}[t]
\centering
\caption{NbR: Fake Injection Success Rates.}
\label{tab:accpoststxt}
\resizebox{\columnwidth}{!}{%
\begin{tabular}{c| c | c | c | c | c| c }\toprule
 \textbf{Strategy} & \textbf{Image} & \textbf{Crime} & \textbf{Safety} & \textbf{Lost Pet} & \multicolumn{1}{|p{1.3cm}|} {\textbf{Unexpected Activity}} & \textbf{Total} \\ \hline 
RSG & n/a  & n/a  & n/a &  n/a &  n/a & 0/100 \\
SGP & n/a  & 9/25 & 6/25 &  10/25   &  1/25  & 23/100 \\
SGA & n/a  & 22/25& 19/25& 16/25   &  9/25  & 66/100 \\ \hline
 SGP & Irrelevant & 9/25 & 6/25 &  10/25  &  1/25 & 23/100 \\
SGA & Irrelevant & 22/25 & \textbf{20/25} & 16/25 &  9/25  & \textbf{67/100} \\ \hline
SGP & Relevant & 9/25& 6/25&  \textbf{16/25}  &  \textbf{2/25} & \textbf{33/100}\\
SGA & Relevant & \textbf{23/25} & \textbf{20/25} & \textbf{25/25}   &  \textbf{12/25}  & \textbf{80/100}\\
\bottomrule
\end{tabular}%
}
\end{table}

\begin{table}[!h]
\centering
\scriptsize
\caption{Examples of accepted fake text posts by NbR.}
\label{tab:accfakex}
\resizebox{\columnwidth}{!}{%
\begin{tabular}{ c| c | c | c } \toprule
\textbf{Model} & \textbf{Category} & \textbf{Fake Title} & \textbf{Fake Description} \\ \hline
SGP & Safety &  \multicolumn{1}{|p{1.3cm}|} {Reports of Fire} & \multicolumn{1}{|p{3cm}} {Reports of Fire. A fire was reported at the Victoria Inn \& Suites on 16th Febuary, 2011. The fire was extinguished with minimal damage to the building. There was no water damage. }\\ \hline

SGP & Lost Pet &  \multicolumn{1}{|p{1cm}|} {Lost My Pet} & \multicolumn{1}{|p{3cm}} {Lost My Pet??? My dog has been missing for a while. We have tried searching for him, but I can't find him. He was a good dog, but he was missing for a long time. I can't find him. Where is he? Is he in the house? }\\\hline

SGA & \multicolumn{1}{|p{1.3cm}|}{Unexpected Activity} & \multicolumn{1}{|p{1.3cm}} {Creeper} & \multicolumn{1}{|p{3cm}} { My neighbor is a creep. He hangs around our yard and keeps looking for food. Last week he came back after we left and stole a can of tomatoes.} \\\hline

SGA & Crime & \multicolumn{1}{|p{1.3cm}|} {Stolen Packages} & \multicolumn{1}{|p{3cm}} {My neighbors kids came in and stole some packages from the front porch. Kids about 12 and under. They were looking for something in a brown bag.} \\ 

\bottomrule
\end{tabular}%
}
\vspace{-10pt}
\end{table}

These results show that \textit{Neighbors by Ring} does check for semantic soundness and also relevance to their categories. The app is also more lenient for ``Lost Pet'' posts, especially if given an image. However, for posts of type ``Crime'', ``Safety'' and ``Unexpected Activity'', the emphasis seems to be on the input text and the information it provides. Nevertheless, we show that using our fake post generation strategies, an adversary can generate multiple posts fulfilling these conditions and effectively perform successful injection attacks. 

\section{Discussion on Countermeasures}
\label{sec:defense}
Majority voting, origin attestation, and reputation schemes can help alleviate improper input injections. However, majority voting depends on the availability of multiple sources of information at any given point in time which is not always true in services with real-time requirements. Origin validation approaches can limit an adversary's ability to scale up the attacks, but they are not effective when used in isolation. \textit{Neighbors by Ring} uses an out-of-band channel for verifying a new user (i.e. email), \textit{Police Detector} assigns user IDs based on unique Android IDs, and \textit{Gas Buddy} and \textit{Google Maps} use client certificates, but they are all bypassable either through technical means as we showed in our work or physical means~\cite{SIMONWEC66:online}. On the other hand, reputation schemes have gained in popularity and found applications in other crowdsourcing domains (e.g. online reviews) but suffer from a cold start problem, which an adversary can leverage to inject only a few but high impact values before they are penalized. Input validation can be a great addition in our defense arsenal which can minimize the adversary's incentive during the cold start period but also throughout the lifetime of a participant account, while rendering the amount of reportings to be potentially checked more tractable. Such strategies are easy to implement, can be immediately deployed with a software update on the server side, and do not assume any capabilities on the participants' devices. 

\begin{table*}[!ht]
\centering
\caption{Example countermeasures and the ensuing reduction in the affected attack surface. $e_1=0.2*350$ and $e_2=0.2*70$. }
\label{tab:defense}
\resizebox{\textwidth}{!}{%
\begin{tabular}{l | l | l | l | l} \toprule
\textbf{App Domain} & \textbf{App Type}   & \textbf{Example Countermeasure} & Function & \textbf{Reduction}  \\ \hline
Strava & Fitness      &  Restrict running distance ($d$) to be at most the world record & $0>d \leq 350m\pm e_1$ & 98.65\%\\
Map My Run & Fitness      &   Restrict running distance ($d$) to be at most the world record & $0>d \leq 350m\pm e_1$ & 99.58\%\\
Fitbit & Fitness      &   Restrict running distance ($d$) to be at most the world record & $0>d \leq 350m\pm e_1$ & 99.58\%\\
Transit & Transportation      &  Enforce bus speed ($v$) according to highway code---70mph in UK Motorways & $0>v\leq 70mph \pm e_2$ & 94.25\%\\
Basket Savings & Pricing     & Use auxiliary data sources to verify price & |aux\_price - reported\_price| < threshold & Varies\\
Police Detector & Location     & Restrict distance ($d()$) between inserted location ($loc(i)$) and the nearest road segment to be within 10m &  $d(loc(i), near(loc(i))) \leq 10m $. & 99.89\% \\
NbR & Safety     &  Use metrics based on user reputation & reputation(user) > threshold & Varies\\

\bottomrule
\end{tabular}
}%
\vspace{-10pt}
\end{table*}

To demonstrate this, we discuss the resounding effect even simple input validation strategies can have in our studied MCSs (see Table~\ref{tab:defense} for a summary). \textit{Strava} (see Section~\ref{sec:fitness}), currently allows a run up to 50 million meters or 31068.56 miles. However, the world record for the longest non-stop run is set by Dean Karnazes to 350 miles. If we were to restrict the input domain of Strava leveraging such domain knowledge (allowing for a 20\% error up to 420 miles which is much higher than the 13m smartphone GPS empirical error~\cite{merry2019smartphone}), this would constitute a reduction of 98.65\% of the original allowed maximum value for distance. The same constraint would reduce \textit{Map My Run} and \textit{Fitbit} input domain to 99.58\% and 95.80\% of their current accepted maximum values. Similarly for \textit{Transit} (Section~\ref{sec:transit}), it is unreasonable to accept bus speed values of 2350km/h ($\approx$1460mph). We could constrain accepted speeds to the maximum speed limit allowed (in the UK buses are allowed up to 70mph on motorways). Allowing for a 20\% error in estimation it would set a limit of 84mph. This constitutes a 94.25\% reduction in the speed allowed, which in turn also greatly limits the effect of the adversary on the expected arrival time of a bus. Both location-based services we analyzed (Section~\ref{sec:location}) could leverage location APIs or public auxiliary data (e.g. the open street map data and the open-source routing machine) to geofence reports within a certain distance from a road segment, or the limits of a target city nearby the reporting device. To demonstrate this we used the \textit{Google Maps API} to restrict accepted GPS values within 10m and 100m of the nearest road segment. We found that out of 2701 values generated with the \textit{CE-2D} strategy, only 3 and 33 of them would be accepted which reduces the accepted locations by 99.89\% and 98.78\% respectively. For pricing services (Section~\ref{sec:prices}, we can restrict the accepted inputs to capture market price fluctuation for the target item. \textit{Basket Savings} can obtain such data through auxiliary data sources~\cite{USDA81:online, ukGovMilk}. Moreover, services like \textit{Basket Savings} can leverage majority voting or employ other verification mechanisms before displaying a value, since grocery item prices do not change frequently~\cite{pesendorfer2002retail}. If real-time distribution is paramount, then the service should employ UI hints to participants with reputation scores of accounts reported a price and abnormal price fluctuation indicators.

In some cases, even if such restrictions are present, a determined adversary can generate fake inputs with a distribution that resembles that of inputs provided by humans, as we demonstrated in the case of \textit{NbR} (Section~\ref{sec:safety}). Detecting fakes is challenging and an open problem. For example, the GPT-2 Output Detector \cite{solaiman2019release} is a model designed specifically to detect fake inputs generated by the GPT-2 model. We tested this model on genuine posts crawled from \textit{NbR} and our fake generated posts. This resulted in a poor precision and recall of 55\% and 53\% respectively. Therefore, the GPT-2 Output Detector cannot be trivially generalized to the target service. While working toward improving our capability to detect such fakes, we should also be displaying visual hints to the users helping them better judge the veracity of a value. For example, every newly created account with no reputation or other maturity and contribution metrics should be marked with a label (e.g. ``NEW') and older accounts should display a reputation score. Accepted values should also be always linked to the account that created them to allow for accountability actions.

\ignore{
\begin{table}[!t]
\centering
\scriptsize
\caption{Example Countermeasures and the ensuing Reduction in the affected attack surface.}
\label{tab:basketcompare}
\resizebox{\columnwidth}{!}{%
\begin{tabular}{l | l | l} \toprule
\textbf{App Type}   & \textbf{Countermeasure} & \textbf{Reduction}  \\ \hline
Fitness      &  Enforce maximum running distance at the world record (420 miles $\pm e$). & 98.65--99.58\%\\
Transit      &  Enforce maximum bus speed according to highway code (70mph $\pm e$ in UK Motorways). & 94.25\%\\
Pricing     & Use auxiliary data sources to verify price. & --\\
Location     & Restrict inserted points within 10m of nearest road segment. & 99.89\% \\
Safety     &  Use metrics based on user reputation. & --\\

\bottomrule
\end{tabular}
}%
\vspace{-10pt}
\end{table}
}

\vspace{-5pt}
\section{Related Work}
\label{sec:related}
Prior work demonstrated data poisoning in crowdsourcing~\cite{naroditskiy2013crowdsourcing, yang2015security, wang2015attacks,carbunar2012you, yao2017automated}. Our work focuses on services interfacing with their users through mobile apps. More related are \cite{polakis2013man, wang2018ghost} but study only specific MCS domains. Polakis et al.~\cite{polakis2013man} demonstrated that two location-based MCSs are vulnerable to fake check-ins. In contrast we observe a more general vulnerability of IIV and use our observation to design a broader study to characterize  the exposure of MCS across domains to improper inputs. Proposed defenses include majority voting~\cite{li2014error,tao2018domain}, reputation systems~\cite{zhang2013secure, wang2016toward} or even trusted sensing~\cite{gilbert2010toward}. However, these are not always applied in MCSs and even if they do their effect is limited when applied in isolation.


Related to our framework, prior works also perform analysis from the perspective of mobile apps but focus on IoT devices rather than MCSs \cite{demetriou2017hanguard, chen2018iotfuzzer}. Others analyze the UI of mobile apps similar to our DEM approach \cite{nan2015uipicker, huang2015supor, hetextexerciser, zhao2020automatic}. In contrast with those works, we do not aim to identify privacy leaks through the UI but instead automate navigation and value fuzzing. Zhao et al.~\cite{zhao2020automatic} also focus on input validation but solely on the mobile app side rather than a remote web service, while others have studied how Android applications' network traffic can be intercepted~\cite{georgiev2012most, sounthiraraj2014smv, 6866553, fahl2012eve}. Furthermore, Zhao et al.~\cite{zhao2019geo} also reverse engineers API calls for Android apps but their work is mostly centered on data leakage vulnerabilities rather than discovering IIV vulnerabilities. We employ similar monitoring strategies which we augment with dynamic instrumentation for bypassing certificate pinning. Lastly, work is undergoing on fake text generation and detection~\cite{yao2017automated,juuti2018stay, adelani2020generating, radford2019language, solaiman2019release, zellers2019defending, gehrmann2019gltr}. Most of them are either not tested in practice or focus on social platforms rather than MCSs. In our work, we leverage such state of the art text generation strategies which we show how they can be combined with spoofed network requests embodied in end-to-end framework for analyzing real-world MCS services exposure to improper input injection attacks.  Lastly, our framework bares similarities with model-based testing~\cite{schieferdecker2012model, utting2012taxonomy, bozic2012model} as it can be seen as a simplistic abstract model for MSCs behavior (the system under test). Building on this modeling is a promising future direction. 


\ignore{
\textbf{crowdsourcing attacks and defenses}. related~\cite{wang2016defending, huang2010you, doan2011crowdsourcing, ali2012crowdits, chatzimilioudis2012crowdsourcing, fuchs2014architecture, aubry2014crowdout, zhang2016privacy, mashhadi2011quality, yang2015security, zhao2019data, eagle2009txteagle}.
Separate attacks and defenses. Why are these different than us (think about their limitations when it comes to being applied in our problem, and their different goals and purpose).

\textbf{analysis through mobile apps} hanguard~\cite{demetriou2017hanguard}, iotfuzzer~\cite{chen2018iotfuzzer}

\textbf{Android UI analysis}. Pluto~\cite{demetriou2016free}, UIPicker~\cite{nan2015uipicker}, SUPOR~\cite{huang2015supor}, TextExerciser~\cite{hetextexerciser}, HiddenBehavior~\cite{zhao2020automatic}.

\textbf{Android static and dynamic code analysis}. Taintdroid, Flowdroid, amandroid. Separate, static vs dynamic. Separate attacks and defenses. Why are these different than our dynamic execution approach. Think about their limitations when it comes to being applied in our problem, and their different goals and purpose.

\textbf{Spoofing network requests}

\textbf{Fake text attacks and defenses}
}

\section{Conclusion}
\label{sec:conclusion}
In this work, we developed a framework for analyzing improper input validation vulnerabilities of mobile crowdsourcing services. We successfully use our framework to analyze 8 high-profile services across 5 application domains and found that they are all severely exposed to improper input injection attacks. Our analysis showed that arbitrary inputs from fake accounts and devices are accepted as genuine, allowing an adversary to fake reports for robberies and gunshots in safety services, to fake fitness activities with supernatural performance, and to manipulate grocery items prices among others. We leverage our insights to propose easy to implement and deploy mitigation strategies based on input domain range and constraint validation and semantic validation, which can greatly reduce the IIV attack surface and contribute to a defense in depth toolkit for mitigating improper input injection attacks attacks.




\begin{acks}
This work was partially supported by the U.S. National Science Foundation (CNS-1956445).
\end{acks}

\bibliographystyle{ACM-Reference-Format}
\bibliography{paper}


\begin{thebibliography}{60}


\ifx \showCODEN    \undefined \def \showCODEN     #1{\unskip}     \fi
\ifx \showDOI      \undefined \def \showDOI       #1{#1}\fi
\ifx \showISBNx    \undefined \def \showISBNx     #1{\unskip}     \fi
\ifx \showISBNxiii \undefined \def \showISBNxiii  #1{\unskip}     \fi
\ifx \showISSN     \undefined \def \showISSN      #1{\unskip}     \fi
\ifx \showLCCN     \undefined \def \showLCCN      #1{\unskip}     \fi
\ifx \shownote     \undefined \def \shownote      #1{#1}          \fi
\ifx \showarticletitle \undefined \def \showarticletitle #1{#1}   \fi
\ifx \showURL      \undefined \def \showURL       {\relax}        \fi
\providecommand\bibfield[2]{#2}
\providecommand\bibinfo[2]{#2}
\providecommand\natexlab[1]{#1}
\providecommand\showeprint[2][]{arXiv:#2}

\bibitem[\protect\citeauthoryear{??}{Fin}{2000}]%
        {FindTheN91:online}
 \bibinfo{year}{2000}\natexlab{}.
\newblock \bibinfo{title}{Gas Buddy}.
\newblock \bibinfo{howpublished}{\url{https://www.gasbuddy.com}}.
\newblock


\bibitem[\protect\citeauthoryear{??}{Ope}{2000}]%
        {OpenCV93:online}
 \bibinfo{year}{2000}\natexlab{}.
\newblock \bibinfo{title}{OpenCV}.
\newblock \bibinfo{howpublished}{\url{https://opencv.org/}}.
\newblock


\bibitem[\protect\citeauthoryear{??}{Map}{2007}]%
        {MapMyRun46:online}
 \bibinfo{year}{2007}\natexlab{}.
\newblock \bibinfo{title}{Map My Run}.
\newblock \bibinfo{howpublished}{\url{https://www.mapmyrun.com/}}.
\newblock


\bibitem[\protect\citeauthoryear{??}{Goo}{2008}]%
        {GoogleMa92:online}
 \bibinfo{year}{2008}\natexlab{}.
\newblock \bibinfo{title}{Google Maps}.
\newblock \bibinfo{howpublished}{\url{https://www.google.com/maps}}.
\newblock


\bibitem[\protect\citeauthoryear{??}{UIA}{2008}]%
        {UIAutoma30:online}
 \bibinfo{year}{2008}\natexlab{}.
\newblock \bibinfo{title}{UI Automator}.
\newblock
  \bibinfo{howpublished}{\url{https://developer.android.com/training/testing/ui-automator.html}}.
\newblock


\bibitem[\protect\citeauthoryear{??}{Fit}{2009}]%
        {FitbitOf86:online}
 \bibinfo{year}{2009}\natexlab{}.
\newblock \bibinfo{title}{Fitbit}.
\newblock \bibinfo{howpublished}{\url{https://www.fitbit.com}}.
\newblock


\bibitem[\protect\citeauthoryear{??}{Str}{2009a}]%
        {StravaSu80:online}
 \bibinfo{year}{2009}\natexlab{a}.
\newblock \bibinfo{title}{Strava}.
\newblock \bibinfo{howpublished}{\url{https://www.strava.com/}}.
\newblock


\bibitem[\protect\citeauthoryear{??}{Str}{2009b}]%
        {StravaLa36:online}
 \bibinfo{year}{2009}\natexlab{b}.
\newblock \bibinfo{title}{Strava Labs}.
\newblock \bibinfo{howpublished}{\url{https://labs.strava.com/}}.
\newblock


\bibitem[\protect\citeauthoryear{??}{Gen}{2011}]%
        {Genymoti8:online}
 \bibinfo{year}{2011}\natexlab{}.
\newblock \bibinfo{title}{Genymotion Android Emulator}.
\newblock \bibinfo{howpublished}{\url{https://www.genymotion.com/}}.
\newblock


\bibitem[\protect\citeauthoryear{??}{Tra}{2012}]%
        {Transit:online}
 \bibinfo{year}{2012}\natexlab{}.
\newblock \bibinfo{title}{Transit}.
\newblock \bibinfo{howpublished}{\url{https://transitapp.com/}}.
\newblock


\bibitem[\protect\citeauthoryear{??}{Pub}{2015}]%
        {PublicTo19:online}
 \bibinfo{year}{2015}\natexlab{}.
\newblock \bibinfo{title}{ToiFi}.
\newblock
  \bibinfo{howpublished}{\url{https://play.google.com/store/apps/details?id=com.apprevelations.indiantoiletfinder}}.
\newblock


\bibitem[\protect\citeauthoryear{??}{Apk}{2016}]%
        {ApktoolA92:online}
 \bibinfo{year}{2016}\natexlab{}.
\newblock \bibinfo{title}{Apktool}.
\newblock \bibinfo{howpublished}{\url{https://ibotpeaches.github.io/Apktool/}}.
\newblock


\bibitem[\protect\citeauthoryear{??}{Bas}{2016}]%
        {BasketSm94:online}
 \bibinfo{year}{2016}\natexlab{}.
\newblock \bibinfo{title}{Basket}.
\newblock \bibinfo{howpublished}{\url{http://basket.com/}}.
\newblock


\bibitem[\protect\citeauthoryear{??}{App}{2017}]%
        {AppiumMo3:online}
 \bibinfo{year}{2017}\natexlab{}.
\newblock \bibinfo{title}{Appium}.
\newblock \bibinfo{howpublished}{\url{http://appium.io/}}.
\newblock


\bibitem[\protect\citeauthoryear{??}{Fri}{2017}]%
        {Frida}
 \bibinfo{year}{2017}\natexlab{}.
\newblock \bibinfo{title}{Frida}.
\newblock \bibinfo{howpublished}{\url{https://frida.re/}}.
\newblock


\bibitem[\protect\citeauthoryear{??}{Nei}{2018}]%
        {Neighbor60:online}
 \bibinfo{year}{2018}\natexlab{}.
\newblock \bibinfo{title}{Neighbors App by Ring}.
\newblock \bibinfo{howpublished}{\url{https://store.ring.com/neighbors}}.
\newblock


\bibitem[\protect\citeauthoryear{??}{Pol}{2018}]%
        {PoliceDe94:online}
 \bibinfo{year}{2018}\natexlab{}.
\newblock \bibinfo{title}{Police Detector (Speed Camera Radar)}.
\newblock
  \bibinfo{howpublished}{\url{https://play.google.com/store/apps/details?id=tat.example.ildar.seer&hl=en_GB}}.
\newblock


\bibitem[\protect\citeauthoryear{??}{Goo}{2019}]%
        {GoogleIm11:online}
 \bibinfo{year}{2019}\natexlab{}.
\newblock \bibinfo{title}{Google Images Download — Google Images Download
  documentation}.
\newblock
  \bibinfo{howpublished}{\url{https://google-images-download.readthedocs.io/en/latest/index.html}}.
\newblock


\bibitem[\protect\citeauthoryear{??}{min}{2019}]%
        {minimaxi63:online}
 \bibinfo{year}{2019}\natexlab{}.
\newblock \bibinfo{title}{minimaxir/gpt-2-simple}.
\newblock
  \bibinfo{howpublished}{\url{https://github.com/minimaxir/gpt-2-simple}}.
\newblock


\bibitem[\protect\citeauthoryear{??}{Clo}{2020}]%
        {CloudNat6:online}
 \bibinfo{year}{2020}\natexlab{}.
\newblock \bibinfo{title}{Google Cloud Natural Language}.
\newblock
  \bibinfo{howpublished}{\url{https://cloud.google.com/natural-language/}}.
\newblock


\bibitem[\protect\citeauthoryear{??}{You}{2020}]%
        {YouVSthe96:online}
 \bibinfo{year}{2020}\natexlab{}.
\newblock \bibinfo{title}{You VS the Year 2020 | MapMyFitness}.
\newblock
  \bibinfo{howpublished}{\url{https://www.mapmyrun.com/challenges/yvsty2020/register}}.
\newblock


\bibitem[\protect\citeauthoryear{??}{mcs}{2021}]%
        {mcs_website}
 \bibinfo{year}{2021}\natexlab{}.
\newblock \bibinfo{title}{Project Website.}
\newblock
  \bibinfo{howpublished}{\url{https://sites.google.com/view/data-poisoning-mcs}}.
\newblock


\bibitem[\protect\citeauthoryear{Adelani, Mai, Fang, Nguyen, Yamagishi, and
  Echizen}{Adelani et~al\mbox{.}}{2020}]%
        {adelani2020generating}
\bibfield{author}{\bibinfo{person}{David~Ifeoluwa Adelani},
  \bibinfo{person}{Haotian Mai}, \bibinfo{person}{Fuming Fang},
  \bibinfo{person}{Huy~H Nguyen}, \bibinfo{person}{Junichi Yamagishi}, {and}
  \bibinfo{person}{Isao Echizen}.} \bibinfo{year}{2020}\natexlab{}.
\newblock \showarticletitle{Generating sentiment-preserving fake online reviews
  using neural language models and their human-and machine-based detection}. In
  \bibinfo{booktitle}{\emph{International Conference on Advanced Information
  Networking and Applications}}. Springer, \bibinfo{pages}{1341--1354}.
\newblock


\bibitem[\protect\citeauthoryear{Bozic and Wotawa}{Bozic and Wotawa}{2012}]%
        {bozic2012model}
\bibfield{author}{\bibinfo{person}{Josip Bozic} {and} \bibinfo{person}{Franz
  Wotawa}.} \bibinfo{year}{2012}\natexlab{}.
\newblock \showarticletitle{Model-based testing-from safety to security}. In
  \bibinfo{booktitle}{\emph{Proceedings of the 9th Workshop on Systems Testing
  and Validation (STV’12)}}. \bibinfo{pages}{9--16}.
\newblock


\bibitem[\protect\citeauthoryear{Carbunar and Potharaju}{Carbunar and
  Potharaju}{2012}]%
        {carbunar2012you}
\bibfield{author}{\bibinfo{person}{Bogdan Carbunar} {and}
  \bibinfo{person}{Rahul Potharaju}.} \bibinfo{year}{2012}\natexlab{}.
\newblock \showarticletitle{You unlocked the mt. everest badge on foursquare!
  countering location fraud in geosocial networks}. In
  \bibinfo{booktitle}{\emph{2012 IEEE 9th International Conference on Mobile
  Ad-Hoc and Sensor Systems (MASS 2012)}}. IEEE, \bibinfo{pages}{182--190}.
\newblock


\bibitem[\protect\citeauthoryear{Chen, Diao, Zhao, Zuo, Lin, Wang, Lau, Sun,
  Yang, and Zhang}{Chen et~al\mbox{.}}{2018}]%
        {chen2018iotfuzzer}
\bibfield{author}{\bibinfo{person}{Jiongyi Chen}, \bibinfo{person}{Wenrui
  Diao}, \bibinfo{person}{Qingchuan Zhao}, \bibinfo{person}{Chaoshun Zuo},
  \bibinfo{person}{Zhiqiang Lin}, \bibinfo{person}{XiaoFeng Wang},
  \bibinfo{person}{Wing~Cheong Lau}, \bibinfo{person}{Menghan Sun},
  \bibinfo{person}{Ronghai Yang}, {and} \bibinfo{person}{Kehuan Zhang}.}
  \bibinfo{year}{2018}\natexlab{}.
\newblock \showarticletitle{IoTFuzzer: Discovering Memory Corruptions in IoT
  Through App-based Fuzzing.}. In \bibinfo{booktitle}{\emph{NDSS}}.
\newblock


\bibitem[\protect\citeauthoryear{Demetriou, Zhang, Lee, Wang, Gunter, Zhou, and
  Grace}{Demetriou et~al\mbox{.}}{2017}]%
        {demetriou2017hanguard}
\bibfield{author}{\bibinfo{person}{Soteris Demetriou}, \bibinfo{person}{Nan
  Zhang}, \bibinfo{person}{Yeonjoon Lee}, \bibinfo{person}{XiaoFeng Wang},
  \bibinfo{person}{Carl~A Gunter}, \bibinfo{person}{Xiaoyong Zhou}, {and}
  \bibinfo{person}{Michael Grace}.} \bibinfo{year}{2017}\natexlab{}.
\newblock \showarticletitle{HanGuard: SDN-driven protection of smart home WiFi
  devices from malicious mobile apps}. In \bibinfo{booktitle}{\emph{Proceedings
  of the 10th ACM Conference on Security and Privacy in Wireless and Mobile
  Networks}}. \bibinfo{pages}{122--133}.
\newblock


\bibitem[\protect\citeauthoryear{Fahl, Harbach, Muders, Baumg{\"a}rtner,
  Freisleben, and Smith}{Fahl et~al\mbox{.}}{2012}]%
        {fahl2012eve}
\bibfield{author}{\bibinfo{person}{Sascha Fahl}, \bibinfo{person}{Marian
  Harbach}, \bibinfo{person}{Thomas Muders}, \bibinfo{person}{Lars
  Baumg{\"a}rtner}, \bibinfo{person}{Bernd Freisleben}, {and}
  \bibinfo{person}{Matthew Smith}.} \bibinfo{year}{2012}\natexlab{}.
\newblock \showarticletitle{Why Eve and Mallory love Android: An analysis of
  Android SSL (in) security}. In \bibinfo{booktitle}{\emph{Proceedings of the
  2012 ACM conference on Computer and communications security}}.
  \bibinfo{pages}{50--61}.
\newblock


\bibitem[\protect\citeauthoryear{Gehrmann, Strobelt, and Rush}{Gehrmann
  et~al\mbox{.}}{2019}]%
        {gehrmann2019gltr}
\bibfield{author}{\bibinfo{person}{Sebastian Gehrmann},
  \bibinfo{person}{Hendrik Strobelt}, {and} \bibinfo{person}{Alexander~M
  Rush}.} \bibinfo{year}{2019}\natexlab{}.
\newblock \showarticletitle{Gltr: Statistical detection and visualization of
  generated text}.
\newblock \bibinfo{journal}{\emph{arXiv preprint arXiv:1906.04043}}
  (\bibinfo{year}{2019}).
\newblock


\bibitem[\protect\citeauthoryear{Georgiev, Iyengar, Jana, Anubhai, Boneh, and
  Shmatikov}{Georgiev et~al\mbox{.}}{2012}]%
        {georgiev2012most}
\bibfield{author}{\bibinfo{person}{Martin Georgiev}, \bibinfo{person}{Subodh
  Iyengar}, \bibinfo{person}{Suman Jana}, \bibinfo{person}{Rishita Anubhai},
  \bibinfo{person}{Dan Boneh}, {and} \bibinfo{person}{Vitaly Shmatikov}.}
  \bibinfo{year}{2012}\natexlab{}.
\newblock \showarticletitle{The most dangerous code in the world: validating
  SSL certificates in non-browser software}. In
  \bibinfo{booktitle}{\emph{Proceedings of the 2012 ACM conference on Computer
  and communications security}}. \bibinfo{pages}{38--49}.
\newblock


\bibitem[\protect\citeauthoryear{Gilbert, Cox, Jung, and Wetherall}{Gilbert
  et~al\mbox{.}}{2010}]%
        {gilbert2010toward}
\bibfield{author}{\bibinfo{person}{Peter Gilbert}, \bibinfo{person}{Landon~P
  Cox}, \bibinfo{person}{Jaeyeon Jung}, {and} \bibinfo{person}{David
  Wetherall}.} \bibinfo{year}{2010}\natexlab{}.
\newblock \showarticletitle{Toward trustworthy mobile sensing}. In
  \bibinfo{booktitle}{\emph{Proceedings of the Eleventh Workshop on Mobile
  Computing Systems \& Applications}}. \bibinfo{pages}{31--36}.
\newblock


\bibitem[\protect\citeauthoryear{gov.uk}{gov.uk}{2020}]%
        {ukGovMilk}
\bibfield{author}{\bibinfo{person}{gov.uk}.} \bibinfo{year}{2020}\natexlab{}.
\newblock \bibinfo{title}{United Kingdom milk prices and composition of milk
  statistics notice (data for April 2020)}.
\newblock
  \bibinfo{howpublished}{\url{https://www.gov.uk/government/publications/uk-milk-prices-and-composition-of-milk/united-kingdom-milk-prices-and-composition-of-milk-statistics-notice-data-for-june-2019}}.
\newblock


\bibitem[\protect\citeauthoryear{He, Zhang, Yang, Cao, Lian, Li, Yang, Zhang,
  Yang, Zhang, et~al\mbox{.}}{He et~al\mbox{.}}{2020}]%
        {hetextexerciser}
\bibfield{author}{\bibinfo{person}{Yuyu He}, \bibinfo{person}{Lei Zhang},
  \bibinfo{person}{Zhemin Yang}, \bibinfo{person}{Yinzhi Cao},
  \bibinfo{person}{Keke Lian}, \bibinfo{person}{Shuai Li}, \bibinfo{person}{Wei
  Yang}, \bibinfo{person}{Zhibo Zhang}, \bibinfo{person}{Min Yang},
  \bibinfo{person}{Yuan Zhang}, {et~al\mbox{.}}}
  \bibinfo{year}{2020}\natexlab{}.
\newblock \showarticletitle{TextExerciser: Feedback-driven Text Input
  Exercising for Android Applications}. In \bibinfo{booktitle}{\emph{2020 IEEE
  Symposium on Security and Privacy}}. IEEE.
\newblock


\bibitem[\protect\citeauthoryear{Huang, Li, Xiao, Wu, Lu, Zhang, and
  Jiang}{Huang et~al\mbox{.}}{2015}]%
        {huang2015supor}
\bibfield{author}{\bibinfo{person}{Jianjun Huang}, \bibinfo{person}{Zhichun
  Li}, \bibinfo{person}{Xusheng Xiao}, \bibinfo{person}{Zhenyu Wu},
  \bibinfo{person}{Kangjie Lu}, \bibinfo{person}{Xiangyu Zhang}, {and}
  \bibinfo{person}{Guofei Jiang}.} \bibinfo{year}{2015}\natexlab{}.
\newblock \showarticletitle{$\{$SUPOR$\}$: Precise and Scalable Sensitive User
  Input Detection for Android Apps}. In \bibinfo{booktitle}{\emph{24th
  $\{$USENIX$\}$ Security Symposium ($\{$USENIX$\}$ Security 15)}}.
  \bibinfo{pages}{977--992}.
\newblock


\bibitem[\protect\citeauthoryear{{Hubbard}, {Weimer}, and {Chen}}{{Hubbard}
  et~al\mbox{.}}{2014}]%
        {6866553}
\bibfield{author}{\bibinfo{person}{J. {Hubbard}}, \bibinfo{person}{K.
  {Weimer}}, {and} \bibinfo{person}{Y. {Chen}}.}
  \bibinfo{year}{2014}\natexlab{}.
\newblock \showarticletitle{A study of SSL Proxy attacks on Android and iOS
  mobile applications}. In \bibinfo{booktitle}{\emph{2014 IEEE 11th Consumer
  Communications and Networking Conference (CCNC)}}. \bibinfo{pages}{86--91}.
\newblock


\bibitem[\protect\citeauthoryear{Juuti, Sun, Mori, and Asokan}{Juuti
  et~al\mbox{.}}{2018}]%
        {juuti2018stay}
\bibfield{author}{\bibinfo{person}{Mika Juuti}, \bibinfo{person}{Bo Sun},
  \bibinfo{person}{Tatsuya Mori}, {and} \bibinfo{person}{N Asokan}.}
  \bibinfo{year}{2018}\natexlab{}.
\newblock \showarticletitle{Stay on-topic: Generating context-specific fake
  restaurant reviews}. In \bibinfo{booktitle}{\emph{European Symposium on
  Research in Computer Security}}. Springer, \bibinfo{pages}{132--151}.
\newblock


\bibitem[\protect\citeauthoryear{Li and Yu}{Li and Yu}{2014}]%
        {li2014error}
\bibfield{author}{\bibinfo{person}{Hongwei Li} {and} \bibinfo{person}{Bin Yu}.}
  \bibinfo{year}{2014}\natexlab{}.
\newblock \showarticletitle{Error rate bounds and iterative weighted majority
  voting for crowdsourcing}.
\newblock \bibinfo{journal}{\emph{arXiv preprint arXiv:1411.4086}}
  (\bibinfo{year}{2014}).
\newblock


\bibitem[\protect\citeauthoryear{Merry and Bettinger}{Merry and
  Bettinger}{2019}]%
        {merry2019smartphone}
\bibfield{author}{\bibinfo{person}{Krista Merry} {and} \bibinfo{person}{Pete
  Bettinger}.} \bibinfo{year}{2019}\natexlab{}.
\newblock \showarticletitle{Smartphone GPS accuracy study in an urban
  environment}.
\newblock \bibinfo{journal}{\emph{PloS one}} \bibinfo{volume}{14},
  \bibinfo{number}{7} (\bibinfo{year}{2019}).
\newblock


\bibitem[\protect\citeauthoryear{Nan, Yang, Yang, Zhou, Gu, and Wang}{Nan
  et~al\mbox{.}}{2015}]%
        {nan2015uipicker}
\bibfield{author}{\bibinfo{person}{Yuhong Nan}, \bibinfo{person}{Min Yang},
  \bibinfo{person}{Zhemin Yang}, \bibinfo{person}{Shunfan Zhou},
  \bibinfo{person}{Guofei Gu}, {and} \bibinfo{person}{XiaoFeng Wang}.}
  \bibinfo{year}{2015}\natexlab{}.
\newblock \showarticletitle{Uipicker: User-input privacy identification in
  mobile applications}. In \bibinfo{booktitle}{\emph{24th $\{$USENIX$\}$
  Security Symposium ($\{$USENIX$\}$ Security 15)}}.
  \bibinfo{pages}{993--1008}.
\newblock


\bibitem[\protect\citeauthoryear{Naroditskiy, Jennings, Van~Hentenryck, and
  Cebrian}{Naroditskiy et~al\mbox{.}}{2013}]%
        {naroditskiy2013crowdsourcing}
\bibfield{author}{\bibinfo{person}{Victor Naroditskiy},
  \bibinfo{person}{Nicholas~R Jennings}, \bibinfo{person}{Pascal
  Van~Hentenryck}, {and} \bibinfo{person}{Manuel Cebrian}.}
  \bibinfo{year}{2013}\natexlab{}.
\newblock \showarticletitle{Crowdsourcing dilemma}.
\newblock \bibinfo{journal}{\emph{arXiv preprint arXiv:1304.3548}}
  (\bibinfo{year}{2013}).
\newblock


\bibitem[\protect\citeauthoryear{Pesendorfer}{Pesendorfer}{2002}]%
        {pesendorfer2002retail}
\bibfield{author}{\bibinfo{person}{Martin Pesendorfer}.}
  \bibinfo{year}{2002}\natexlab{}.
\newblock \showarticletitle{Retail sales: A study of pricing behavior in
  supermarkets}.
\newblock \bibinfo{journal}{\emph{The Journal of Business}}
  \bibinfo{volume}{75}, \bibinfo{number}{1} (\bibinfo{year}{2002}),
  \bibinfo{pages}{33--66}.
\newblock


\bibitem[\protect\citeauthoryear{Polakis, Volanis, Athanasopoulos, and
  Markatos}{Polakis et~al\mbox{.}}{2013}]%
        {polakis2013man}
\bibfield{author}{\bibinfo{person}{Iasonas Polakis}, \bibinfo{person}{Stamatis
  Volanis}, \bibinfo{person}{Elias Athanasopoulos}, {and}
  \bibinfo{person}{Evangelos~P Markatos}.} \bibinfo{year}{2013}\natexlab{}.
\newblock \showarticletitle{The man who was there: validating check-ins in
  location-based services}. In \bibinfo{booktitle}{\emph{Proceedings of the
  29th Annual Computer Security Applications Conference}}.
  \bibinfo{pages}{19--28}.
\newblock


\bibitem[\protect\citeauthoryear{Radford, Wu, Child, Luan, Amodei, and
  Sutskever}{Radford et~al\mbox{.}}{2019}]%
        {radford2019language}
\bibfield{author}{\bibinfo{person}{Alec Radford}, \bibinfo{person}{Jeffrey Wu},
  \bibinfo{person}{Rewon Child}, \bibinfo{person}{David Luan},
  \bibinfo{person}{Dario Amodei}, {and} \bibinfo{person}{Ilya Sutskever}.}
  \bibinfo{year}{2019}\natexlab{}.
\newblock \showarticletitle{Language models are unsupervised multitask
  learners}.
\newblock \bibinfo{journal}{\emph{OpenAI Blog}} \bibinfo{volume}{1},
  \bibinfo{number}{8} (\bibinfo{year}{2019}), \bibinfo{pages}{9}.
\newblock


\bibitem[\protect\citeauthoryear{Schieferdecker}{Schieferdecker}{2012}]%
        {schieferdecker2012model}
\bibfield{author}{\bibinfo{person}{Ina Schieferdecker}.}
  \bibinfo{year}{2012}\natexlab{}.
\newblock \showarticletitle{Model-Based Fuzz Testing}. In
  \bibinfo{booktitle}{\emph{2012 IEEE Fifth International Conference on
  Software Testing, Verification and Validation}}. IEEE,
  \bibinfo{pages}{814--814}.
\newblock


\bibitem[\protect\citeauthoryear{Simon}{Simon}{2020}]%
        {SIMONWEC66:online}
\bibfield{author}{\bibinfo{person}{Weckert Simon}.}
  \bibinfo{year}{2020}\natexlab{}.
\newblock \bibinfo{title}{Google Maps Hacks}.
\newblock
  \bibinfo{howpublished}{\url{http://www.simonweckert.com/googlemapshacks.html}}.
\newblock


\bibitem[\protect\citeauthoryear{Solaiman, Brundage, Clark, Askell,
  Herbert-Voss, Wu, Radford, and Wang}{Solaiman et~al\mbox{.}}{2019}]%
        {solaiman2019release}
\bibfield{author}{\bibinfo{person}{Irene Solaiman}, \bibinfo{person}{Miles
  Brundage}, \bibinfo{person}{Jack Clark}, \bibinfo{person}{Amanda Askell},
  \bibinfo{person}{Ariel Herbert-Voss}, \bibinfo{person}{Jeff Wu},
  \bibinfo{person}{Alec Radford}, {and} \bibinfo{person}{Jasmine Wang}.}
  \bibinfo{year}{2019}\natexlab{}.
\newblock \showarticletitle{Release strategies and the social impacts of
  language models}.
\newblock \bibinfo{journal}{\emph{arXiv preprint arXiv:1908.09203}}
  (\bibinfo{year}{2019}).
\newblock


\bibitem[\protect\citeauthoryear{Sounthiraraj, Sahs, Greenwood, Lin, and
  Khan}{Sounthiraraj et~al\mbox{.}}{2014}]%
        {sounthiraraj2014smv}
\bibfield{author}{\bibinfo{person}{David Sounthiraraj}, \bibinfo{person}{Justin
  Sahs}, \bibinfo{person}{Garret Greenwood}, \bibinfo{person}{Zhiqiang Lin},
  {and} \bibinfo{person}{Latifur Khan}.} \bibinfo{year}{2014}\natexlab{}.
\newblock \showarticletitle{Smv-hunter: Large scale, automated detection of
  ssl/tls man-in-the-middle vulnerabilities in android apps}. In
  \bibinfo{booktitle}{\emph{In Proceedings of the 21st Annual Network and
  Distributed System Security Symposium (NDSS’14}}. Citeseer.
\newblock


\bibitem[\protect\citeauthoryear{Tao, Cheng, Yu, Yue, and Wang}{Tao
  et~al\mbox{.}}{2018}]%
        {tao2018domain}
\bibfield{author}{\bibinfo{person}{Dapeng Tao}, \bibinfo{person}{Jun Cheng},
  \bibinfo{person}{Zhengtao Yu}, \bibinfo{person}{Kun Yue}, {and}
  \bibinfo{person}{Lizhen Wang}.} \bibinfo{year}{2018}\natexlab{}.
\newblock \showarticletitle{Domain-weighted majority voting for crowdsourcing}.
\newblock \bibinfo{journal}{\emph{IEEE transactions on neural networks and
  learning systems}} \bibinfo{volume}{30}, \bibinfo{number}{1}
  (\bibinfo{year}{2018}), \bibinfo{pages}{163--174}.
\newblock


\bibitem[\protect\citeauthoryear{usda.gov}{usda.gov}{2019}]%
        {USDA81:online}
\bibfield{author}{\bibinfo{person}{usda.gov}.} \bibinfo{year}{2019}\natexlab{}.
\newblock \bibinfo{title}{Price Spreads from Farm to Consumer}.
\newblock
  \bibinfo{howpublished}{\url{https://www.ers.usda.gov/data-products/price-spreads-from-farm-to-consumer/}}.
\newblock


\bibitem[\protect\citeauthoryear{Utting, Pretschner, and Legeard}{Utting
  et~al\mbox{.}}{2012}]%
        {utting2012taxonomy}
\bibfield{author}{\bibinfo{person}{Mark Utting}, \bibinfo{person}{Alexander
  Pretschner}, {and} \bibinfo{person}{Bruno Legeard}.}
  \bibinfo{year}{2012}\natexlab{}.
\newblock \showarticletitle{A taxonomy of model-based testing approaches}.
\newblock \bibinfo{journal}{\emph{Software testing, verification and
  reliability}} \bibinfo{volume}{22}, \bibinfo{number}{5}
  (\bibinfo{year}{2012}), \bibinfo{pages}{297--312}.
\newblock


\bibitem[\protect\citeauthoryear{Wang, Wang, Wang, Nika, Liu, Zheng, and
  Zhao}{Wang et~al\mbox{.}}{2015}]%
        {wang2015attacks}
\bibfield{author}{\bibinfo{person}{Gang Wang}, \bibinfo{person}{Bolun Wang},
  \bibinfo{person}{Tianyi Wang}, \bibinfo{person}{Ana Nika},
  \bibinfo{person}{Bingzhe Liu}, \bibinfo{person}{Haitao Zheng}, {and}
  \bibinfo{person}{Ben~Y Zhao}.} \bibinfo{year}{2015}\natexlab{}.
\newblock \showarticletitle{Attacks and defenses in crowdsourced mapping
  services}.
\newblock \bibinfo{journal}{\emph{CoRR, abs/1508.00837}}
  (\bibinfo{year}{2015}).
\newblock


\bibitem[\protect\citeauthoryear{Wang, Wang, Wang, Nika, Zheng, and Zhao}{Wang
  et~al\mbox{.}}{2016b}]%
        {wang2016defending}
\bibfield{author}{\bibinfo{person}{Gang Wang}, \bibinfo{person}{Bolun Wang},
  \bibinfo{person}{Tianyi Wang}, \bibinfo{person}{Ana Nika},
  \bibinfo{person}{Haitao Zheng}, {and} \bibinfo{person}{Ben~Y Zhao}.}
  \bibinfo{year}{2016}\natexlab{b}.
\newblock \showarticletitle{Defending against sybil devices in crowdsourced
  mapping services}. In \bibinfo{booktitle}{\emph{Proceedings of the 14th
  Annual International Conference on Mobile Systems, Applications, and
  Services}}. \bibinfo{pages}{179--191}.
\newblock


\bibitem[\protect\citeauthoryear{Wang, Wang, Wang, Nika, Zheng, and Zhao}{Wang
  et~al\mbox{.}}{2018}]%
        {wang2018ghost}
\bibfield{author}{\bibinfo{person}{Gang Wang}, \bibinfo{person}{Bolun Wang},
  \bibinfo{person}{Tianyi Wang}, \bibinfo{person}{Ana Nika},
  \bibinfo{person}{Haitao Zheng}, {and} \bibinfo{person}{Ben~Y Zhao}.}
  \bibinfo{year}{2018}\natexlab{}.
\newblock \showarticletitle{Ghost riders: Sybil attacks on crowdsourced mobile
  mapping services}.
\newblock \bibinfo{journal}{\emph{IEEE/ACM transactions on networking}}
  \bibinfo{volume}{26}, \bibinfo{number}{3} (\bibinfo{year}{2018}),
  \bibinfo{pages}{1123--1136}.
\newblock


\bibitem[\protect\citeauthoryear{Wang, Qi, Shu, Deng, and Rodrigues}{Wang
  et~al\mbox{.}}{2016a}]%
        {wang2016toward}
\bibfield{author}{\bibinfo{person}{Kun Wang}, \bibinfo{person}{Xin Qi},
  \bibinfo{person}{Lei Shu}, \bibinfo{person}{Der-jiunn Deng}, {and}
  \bibinfo{person}{Joel~JPC Rodrigues}.} \bibinfo{year}{2016}\natexlab{a}.
\newblock \showarticletitle{Toward trustworthy crowdsourcing in the social
  internet of things}.
\newblock \bibinfo{journal}{\emph{IEEE Wireless Communications}}
  \bibinfo{volume}{23}, \bibinfo{number}{5} (\bibinfo{year}{2016}),
  \bibinfo{pages}{30--36}.
\newblock


\bibitem[\protect\citeauthoryear{Yang, Zhang, Ren, and Shen}{Yang
  et~al\mbox{.}}{2015}]%
        {yang2015security}
\bibfield{author}{\bibinfo{person}{Kan Yang}, \bibinfo{person}{Kuan Zhang},
  \bibinfo{person}{Ju Ren}, {and} \bibinfo{person}{Xuemin Shen}.}
  \bibinfo{year}{2015}\natexlab{}.
\newblock \showarticletitle{Security and privacy in mobile crowdsourcing
  networks: challenges and opportunities}.
\newblock \bibinfo{journal}{\emph{IEEE communications magazine}}
  \bibinfo{volume}{53}, \bibinfo{number}{8} (\bibinfo{year}{2015}),
  \bibinfo{pages}{75--81}.
\newblock


\bibitem[\protect\citeauthoryear{Yao, Viswanath, Cryan, Zheng, and Zhao}{Yao
  et~al\mbox{.}}{2017}]%
        {yao2017automated}
\bibfield{author}{\bibinfo{person}{Yuanshun Yao}, \bibinfo{person}{Bimal
  Viswanath}, \bibinfo{person}{Jenna Cryan}, \bibinfo{person}{Haitao Zheng},
  {and} \bibinfo{person}{Ben~Y Zhao}.} \bibinfo{year}{2017}\natexlab{}.
\newblock \showarticletitle{Automated crowdturfing attacks and defenses in
  online review systems}. In \bibinfo{booktitle}{\emph{Proceedings of the 2017
  ACM SIGSAC Conference on Computer and Communications Security}}.
  \bibinfo{pages}{1143--1158}.
\newblock


\bibitem[\protect\citeauthoryear{Zellers, Holtzman, Rashkin, Bisk, Farhadi,
  Roesner, and Choi}{Zellers et~al\mbox{.}}{2019}]%
        {zellers2019defending}
\bibfield{author}{\bibinfo{person}{Rowan Zellers}, \bibinfo{person}{Ari
  Holtzman}, \bibinfo{person}{Hannah Rashkin}, \bibinfo{person}{Yonatan Bisk},
  \bibinfo{person}{Ali Farhadi}, \bibinfo{person}{Franziska Roesner}, {and}
  \bibinfo{person}{Yejin Choi}.} \bibinfo{year}{2019}\natexlab{}.
\newblock \showarticletitle{Defending against neural fake news}. In
  \bibinfo{booktitle}{\emph{Advances in Neural Information Processing
  Systems}}. \bibinfo{pages}{9051--9062}.
\newblock


\bibitem[\protect\citeauthoryear{Zhang, Zhang, Zhang, and Zhang}{Zhang
  et~al\mbox{.}}{2013}]%
        {zhang2013secure}
\bibfield{author}{\bibinfo{person}{Rui Zhang}, \bibinfo{person}{Jinxue Zhang},
  \bibinfo{person}{Yanchao Zhang}, {and} \bibinfo{person}{Chi Zhang}.}
  \bibinfo{year}{2013}\natexlab{}.
\newblock \showarticletitle{Secure crowdsourcing-based cooperative pectrum
  sensing}. In \bibinfo{booktitle}{\emph{2013 Proceedings IEEE INFOCOM}}. IEEE,
  \bibinfo{pages}{2526--2534}.
\newblock


\bibitem[\protect\citeauthoryear{Zhao, Zuo, Brendan, Pellegrino, and Lin}{Zhao
  et~al\mbox{.}}{2020}]%
        {zhao2020automatic}
\bibfield{author}{\bibinfo{person}{Qingchuan Zhao}, \bibinfo{person}{Chaoshun
  Zuo}, \bibinfo{person}{Dolan-Gavitt Brendan}, \bibinfo{person}{Giancarlo
  Pellegrino}, {and} \bibinfo{person}{Zhiqiang Lin}.}
  \bibinfo{year}{2020}\natexlab{}.
\newblock \showarticletitle{Automatic Uncovering of Hidden Behaviors From Input
  Validation in Mobile Apps}. In \bibinfo{booktitle}{\emph{2020 IEEE Symposium
  on Security and Privacy}}. IEEE.
\newblock


\bibitem[\protect\citeauthoryear{Zhao, Zuo, Pellegrino, and Zhiqiang}{Zhao
  et~al\mbox{.}}{2019}]%
        {zhao2019geo}
\bibfield{author}{\bibinfo{person}{Qingchuan Zhao}, \bibinfo{person}{Chaoshun
  Zuo}, \bibinfo{person}{Giancarlo Pellegrino}, {and} \bibinfo{person}{Li
  Zhiqiang}.} \bibinfo{year}{2019}\natexlab{}.
\newblock \showarticletitle{Geo-locating Drivers: A Study of Sensitive Data
  Leakage in Ride-Hailing Services.}. In \bibinfo{booktitle}{\emph{Annual
  Network and Distributed System Security symposium, February 2019 (NDSS
  2019)}}.
\newblock


\end{thebibliography}

\appendix
\label{appendix}

\section{Other Services}
\label{app:other_services}
\vspace{0pt}\noindent\textbf{Fitbit}~\cite{FitbitOf86:online} is a fitness service. Its Android app is installed more than 10,000,000 times. It allows its users to record walks, hikes and runs, share their achievements with others, participate in challenges and earn badges as rewards for their activities. An adversary faking activities can earn badges, win challenges and rewards. To demonstrate Fitbit's susceptibility to improper input injections we spoof network requests to the Fitbit service. We were successful in posting activities spoofing a real device and user account. To verify the injection, we use the response of the spoofed requests and we also verify manually on a real device in parallel for some randomly chosen requests. To characterize the IIV attack surface of Fitbit we follow the \textit{NVE} approach (see Section~\ref{sec:framework}), to study the successful injection boundaries for an activity's distance and duration. We repeat for three activities: walk, hike and run. We found that no negative or zero values for distance or duration are allowed. Distance can be added from 1km to 1609.344km (equivalent to 10,000 miles). For duration we observe that there is no input data type positive range restriction as we could inject durations from 1 second to 2,147,483,647 (which is the maximum positive value for a 32-bit signed binary integer, or $2^{31}-1$).

\vspace{5pt}\noindent\textbf{Run with Map My Run}~\cite{MapMyRun46:online} is a fitness tracking service whose Android app is also installed more than 10,000,000 times. Like other fitness services, it allows its users to track and share activities, participate in challenges, and earn rewards. Its challenges can be sponsored and result in real in-kind rewards. For example, the \textit{You VS the Year 2020} challenge asks users to \textit{``Cover 1,020KM in 2020 and be eligible for exclusive prizes from Under Armour''}~\cite{YouVSthe96:online}. Like before, we were able to reverse engineer the service's API and successfully spoof network requests to the service, which appear to come from a real device and user, and observe the result of the injection. To characterize the adversary's reach, we perform experiments on the distance and duration of a running activity using the \textit{NVE} strategy.  


We found that the min duration that can be added is 0. For the max value, we found that the remote service accepts an arbitrarily large value for the duration, but this is always rounded within the constraints of a single day. That is, if value $x$ in seconds is stored on the server-side, the client always renders $x\mod 86400$. This is alarming because any input validation happens on the client-side while the values are stored non validated on the server. Thus, even though we do not know whether the stored values are sanitized or not on the server-side\ignore{before being used for statistical aggregation or for informing learning algorithms}, this would require a lot of ad-hoc validation checks whenever the value is to be used, which is a bad prone to errors design. In terms of distance, we found that the min distance accepted is 0. However, the max allowed distance for an activity can be arbitrarily large. We were able to increase it up to 100051.4 miles, which is approximately 4 times the earth's perimeter.

\section{Ethical Considerations}
\label{sec:ethical_considerations_appendix}

\subsection{Details on IRB Approval}

We applied for an IRB at the University of Southern California which can be verified by study-iD:IIR00003094. The following is an excerpt from the response we received: “Therefore, this study is considered Not Human Subjects Research* and is not subject to 45 CFR 46 regulations, including informed consent requirements or further IRB review.”. Our study do not qualify for IRB approval because we do not collect data through interaction with humans and do not  collect personally identifiable information. 
 
Further details from the IRB response email are provided below: 

\begin{shaded}
\vspace{3pt}\noindent The University Park Institutional Review Board (UPIRB) designee reviewed the information you submitted pertaining to your study and has determined on 12/10/2019 that the project does not qualify as Human Subjects Research.

\vspace{3pt}\noindent From 45 CFR 46.102, The Federal Regulations on Human Subjects Research is as follows:  

\vspace{3pt}\noindent Human Subject: A living individual about whom an investigator (whether professional or student) conducting research obtains data through intervention or interaction with the individual, or identifiable private information. 

\vspace{3pt}\noindent Research: A systematic investigation, including research development, testing, and evaluation, designed to develop or contribute to generalizable knowledge.
\end{shaded}

Nonetheless, we integrated various measures in the design of our experiments' to protect users and apps from harm. Next we elaborate on some of those measures.

\subsection{Basket Savings}
The experiments were conducted after midnight when the grocery stores are closed. Item prices were reverted to their original value right after each experiment.


\subsection{Fitbit}
The added fake activities were deleted right after the experiments. The fake accounts did not add any real people as friends on the app to minimize interaction with actual users on the app and we did not participate in any real challenges to win rewards from a leaderboard. 

\subsection{Police Detector}
The vast majority of the inserted POIs were positioned in the ocean or away from human populations, therefore it is unlikely that people were misled by them. We also deleted all POIs right after their insertion to ensure that fake POIs could only be visible for a couple of seconds. 

\subsection{Transit}
Our experiments were performed targeting the same bus route (18km) in a rural area. To minimize influence, we performed our experiments during off-peak hours. \textit{Transit} displays the number of users viewing one's contribution. In our experiments, we verified that we only affected our observer devices and no actual user was affected. 

\subsection{Neighbors by Ring}
We set our location in Death Valley to limit the possible number of inhabitants and checked that there were no activity within the app at and around that location. We also removed the accepted posts within 8 min of being published. Furthermore, we only interacted with the algorithms and the service itself and collected data that is openly available and non-identifiable.

\subsection{Strava}
We did not participate in any real challenges to win real rewards from leaderboards. Furthermore, we did not add any friends on the app which means no other user got notified or could view our fake activities. The added fake accounts and fake activities were deleted right after the experiments.
 
\subsection{Map My Run}
We did not participate in any real challenges to win real rewards from leaderboards. Furthermore, we did not add any friends on the app which means no other user got notified or  could view our fake activities. The added fake accounts and fake activities were deleted right after the experiments.

 
\subsection{Toifi (Toilet Finder)}
The majority of the inserted POIs were located in the ocean or away from human populations. All POIs were deleted the POIs right after inserting which means they were only visible for a couple of seconds at most. 

\subsection{Google Maps}

The experiments were conducted within the university campus and after midnight to minimize the effect on users. For the experiments, we were physically present to ensure that there were no people present at the time and could abort the experiment otherwise. 
 
\subsection{Gas Buddy}
We reverted the prices back to normal right after changing them, which means they were visible on the app for a couple of seconds. All experiments took place after midnight to further minimize potential exposure of users to the fake prices.

\end{document}